\documentclass[preprint]{aastex}
\slugcomment{submitted to {\it The Astronomical Journal}}
\begin{document}
\title{Lithium in the Intermediate-Age Open Cluster, NGC 3680}
\author{Barbara J. Anthony-Twarog$^1$, Constantine P. Deliyannis$^{2,3}$, Bruce A. Twarog$^{1,3}$, Kevin V. Croxall$^2$, Jeffrey D. Cummings$^2$}
\affil{$^1$Department of Physics and Astronomy, University of Kansas, Lawrence, KS 66045-7582}
\affil{$^2$Department of Astronomy, Indiana University, Bloomington, IN 47405-7105}
\affil{Electronic mail: bjat@ku.edu, con@astro.indiana.edu, btwarog@ku.edu, kcroxall@indiana.edu, jdcummi@astro.indiana.edu}
\altaffiltext{3}{Visiting Astronomer, Cerro Tololo Inter-American Observatory. CTIO is operated by AURA, Inc., under contract to the National Science Foundation.}

\begin{abstract}
High-dispersion spectra centered on the Li 6708 \AA\ line have been obtained for 70 potential members of the 
intermediate-age open cluster NGC 3680, with an emphasis on stars in the turnoff region of the cluster CMD. A measurable Li 
abundance has been derived for 53 stars, 39 of which have radial velocities and proper motions
consistent with cluster membership. After being transferred to common temperature and abundance scales, previous Li estimates 
have been combined to generate a sample of 49 members, 40 of which bracket the cluster Li-dip. Spectroscopic elemental 
analysis of 8 giants and 5 turnoff stars produces [Fe/H] = -0.17 $\pm$ 0.07 (sd) and -0.07 $\pm$ 0.02 (sd), respectively. We also report measurements of Ca, Si and Ni which are consistent with scaled-solar ratios within the errors. Adopting [Fe/H] = -0.08 (\S3.6), Y$^2$ isochrone comparisons 
lead to an age of 1.75 $\pm$ 0.1 Gyr and an apparent modulus of $(m-M)$ = 10.30 $\pm$ 0.15 for the cluster, placing the center of the Li-dip at 
1.35 $\pm$ 0.03 $M_{\sun}$. Among the giants, five of nine cluster members are now known to have measurable Li with A(Li) near 1.0. 
A combined sample of dwarfs in the Hyades and Praesepe is used to delineate the Li-dip profile at 0.7 Gyr 
and [Fe/H] = +0.15, establishing its center at 1.42 $\pm$ 0.02 $M_{\sun}$ and noting the possible existence of secondary dip on its red boundary. When evolved to the typical age of the clusters 
NGC 752 (age = 1.45 Gyr, $(m-M)$ = 8.4), IC 4651 (age = 1.5 Gyr, $(m-M)$ = 10.4), and NGC 3680, the Hyades/Praesepe Li-dip profile 
reproduces the observed morphology of the combined Li-dip within the CMD's of the intermediate-age clusters while implying a 
metallicity dependence for the central mass of the Li-dip given by M/M$_{\sun}$ = 1.38 $\pm$0.04 + 0.4 $\pm$0.2 [Fe/H]. The implications of the 
similarity of the Li-dichotomy among giants in NGC 752 and IC 4651 and the disagreement with the pattern among NGC 3680 giants are discussed.
\end{abstract}


\section{INTRODUCTION}
Open clusters are studied for a variety of reasons, ranging from their intended use as probes of the structural 
and chemical evolution of the Galaxy \citep{fri02, yon05, ses08} to delineation of the finer details of stellar 
structure and evolution \citep{mei02, van04}. While lithium is an element with clear cosmological origins and 
ramifications, studies of lithium in open clusters predominantly emphasize its role within stellar evolution, though 
insights gained from these investigations could have an indirect impact upon our understanding of the cosmological 
parameters \citep{sr05, kor06, cd00}. Lithium, boron and beryllium are relatively fragile nuclei. Attrition of lithium 
from the presumed initial value of A(Li) = 3.30 implied by meteoritic abundances \citep{ag89} and very young star clusters, is expected in the 
internal layers of stars or protostars where the temperatures exceed $\sim 2.5 \times 10^6$ K, above which Li burns 
via proton capture to $^4$He. Over a star's lifetime, the integrated effect of any structural features that put 
surface layers in contact with the deeper layers at or above this temperature will be expressed in the depletion of 
surface Li. The surface Li abundance can also be enhanced if radiative acceleration pushes it into the surface convection zone \citep{pm93,caj} or if a deepening convection zone dredges up previously diffused Li \citep{ddk,sd}.

Standard\footnote[1]{By 'standard' we mean spherically symmetric models that ignore effects due to diffusion, rotation, mass loss and magnetic fields.}
stellar evolution theory (SSET) predicts the attrition of Li as a function of stellar mass, age, and composition.  Low mass pre-main sequence models evolving down the Hayashi track develop high enough temperatures at the base of the surface convection zone (SCZ) so that energetic protons destroy Li there, thereby depleting the surface Li abundance.  Lower stellar mass implies a deeper and longer-lived SCZ, which implies more Li depletion; there is little Li depletion for models reaching the ZAMS as A, F, and G dwarfs, while Li depletion is more severe for K and cooler spectral types (e.g. \citet{pin}).  Li depletion ceases as the SCZ recedes toward the surface and its base cools.  Independently of the details of the (standard) input physics, SSET {\it robustly} predicts that no further Li depletion occurs during the MS for AFG dwarfs (e.g. \citep{pm89, ddk, pdk, sfr, cdp}) whereas KM dwarfs can continue to deplete Li as they evolve beyond the ZAMS.  The predicted Li-T$_{e}$ relation is modified very slightly as models evolve to slightly higher T$_{e}$ over time.  Higher-metallicity models have deeper convection and deplete more Li.

These predictions are qualitatively consistent with observations of main sequence stars in younger open clusters. 
Observations of stars in young clusters like the Pleiades \citep{so93} do confirm the general ZAMS Li-T$_{e}$ relation resulting 
from pre-main-sequence evolution. However, stars in real clusters exhibit major departures from the theoretical expectations. 
A significant dispersion in the Li abundance is found at any given stellar mass or 
surface temperature, implying that factors beyond the star's basic structural parameters of mass, chemical composition and 
age are relevant \citep{ki00}. An additional clue is posed by the observation that short-period, tidally-locked binaries 
retain their high initial surface Li, an association that highlights rotation and/or stellar rotational history as additional 
factors \citep{del94}, in agreement with studies of both field stars \citep{ran99, che01, do03, ma03, jas06} and 
clusters \citep{tho93, jo99, pas04}.

Among older open clusters, depletion of Li in low mass stars as a function of stellar mass exceeds SSET predictions 
by amounts that grow increasingly disparate with age. At ages between 0.1 to 0.2 Gyr, a feature strikingly at odds with SSET 
begins to appear in plots of Li abundance versus stellar surface temperature or mass \citep{ascon}. By the age of the Hyades (0.65 Gyrs), 
stars with surface temperatures 
of $6600 \pm 200$ K, consistent with main sequence masses near 1.4 $M_{\sun}$, show abrupt depletions in Li of up to 
2 orders of magnitude, defining the Li or Boesgaard-Tripicco dip \citep{bt86}. Such stars would not be expected to have 
convective envelopes; no obvious physical mechanism exists to put the surface layer in contact with the hotter, deeper 
layers. As emphasized by the many papers evaluating the properties relevant to this distinct 
feature \citep{pas01, che01, boe02, ma03, pas04, ascon}, mapping the depth of the dip and its range in 
temperature and mass as a function of metallicity and age is critical to any attempt to decipher the underlying physical mechanism 
that produces it \citep{cha05}. Moreover, among clusters in the 3 to 5 Gyr range, the stars leaving the main sequence 
for the subgiant and giant branch emerge from the Li-dip with inherently lower Li that should be driven even lower by post-main 
sequence evolution. Discovery of any evolved stars with measurable Li in clusters of this age range should be a clear 
indicator of anomalous evolution, mass transfer, and/or deep mixing of newly processed Li to the surface of stars on the giant 
branch \citep{sb95, pas01, pas04, att04}. Finally, the common age of cluster dwarfs offers the opportunity of defining reliably 
the time evolution of the Li depletion among the cooler stars beyond the Li-dip if significant samples with sufficient 
spectroscopic accuracy can be compiled over the age range of interest \citep{sr05, ran06, pri07}.

The focus of this investigation is the intermediate-age open cluster, NGC 3680. It was selected as a prelude to the analysis 
of a significantly more challenging but exceptional cluster, NGC 6253, for several reasons.  As discussed in more detail in Sec. 3.3, some Li 
data for NGC 3680 stars from the turnoff through the giant branch do exist \citep{pas01} (PRP), though the sample at and below the turnoff is 
by no means complete; we hoped to broaden the characterization of the Li-dip profile by expanding the sample with comparably high signal-to-noise spectra.  At the same time, we wanted adequate overlap with previous work, not just so that conclusions might be strengthened by combined results for
common stars in NGC 3680 but also to assist in validating our results for NGC 6253, based on spectra obtained with the same equipment and similarly reduced. 
Third, while photometric metallicity estimates are available 
for the turnoff stars and giants, as well as moderate-dispersion spectroscopic abundances for a few giants, metal abundances based on 
high-dispersion spectroscopy have been obtained for only a few cluster giants \citep{pas01, smi08, san08} and two dwarfs \citep{pac08, san08}, one of which might not be a member.  Obtaining sufficiently high S/N spectra for NGC 3680's dwarf stars does not present an insuperable challenge, but the task is complicated by the severe paucity of single dwarfs in the cluster.  

Broader astrophysical circumstances make NGC 3680 an intriguing target for observation. It is one of a trio of well-studied 
clusters of similar age: NGC 752 is closer than NGC 3680 but of comparable metallicity, while IC 4651 is more richly populated, more 
distant and definitely more metal-rich than either cluster, as confirmed by both photometry \citep{att87, nis88, att00, mei02} and 
spectroscopy \citep{pas04, car04, pac08}. However, the modest metallicity range of $\sim$0.3 dex among the trio brackets the range 
occupied by the large majority of open clusters of any age within 1.5 kpc of the Sun \citep{taat}. Along with NGC 752 \citep{tw83, dan94}, 
the color-magnitude diagram (CMD) of NGC 3680 exhibits a {\it bimodal} main sequence \citep{nis88, atts}, i.e., a secondary sequence 
displaced to the red of the comparably populated main sequence, the byproduct of a rich binary population \citep{dan94, naa97}. Having 
collated photometry, proper motions and radial velocities, \citet{naa97} derived a binary fraction of 57\% among $\sim$100 members of NGC 
3680 with complete data. 

NGC 3680, like IC 4651 and NGC 752, appears to have a poor to modest population below $1 M_{\sun}$ \citep{att91, haw99}. 
Based on the extant stellar population, \citet{naa97} estimate that the cluster mass might have been ten times greater and 
the stellar population 30 times more numerous before its apparent evaporation.

The intermediate age of $\sim 1.7$ Gyr for the three clusters places their turnoff stars squarely in the region of the Li-dip, 
but young enough that stars higher in mass than the dip should still be present near the turnoff region. The turnoff morphology 
also implies a turnoff mass that simultaneously corresponds to the mass for which convective core overshoot should play a 
conspicuous role \citep{mp88, atts, acn90, dan94, kp97}.  Confirmation of these characteristics has been based 
on exquisite information about proper-motion and radial-velocity memberships and binarity \citep{kp95, merm95, naa96}.  
Once the numerous binaries are identified and removed from the CMD, the location of the single-star 
main sequence most resembles stellar models incorporating significant convective overshoot \citep{naa97}. 

The layout of the remainder of the paper is as follows: Sec. 2 details the spectroscopic observations and their reduction to produce
radial velocities used in conjunction with proper motions to identify cluster members. Sec. 3 presents the fundamental parameters that 
play a role in the analysis of the spectra, as well as the derivation and interpretation of the abundances. Sec. 4 discusses the Li 
abundance pattern in NGC 3680, NGC 752, IC 4651, the Hyades, and Praesepe. The composite structure of the Li-dip and its possible 
evolution are revisited in Sec. 5 and our conclusions are summarized in Sec. 6.

\section{SPECTROSCOPIC DATA}
\subsection{Observations and Reductions}
Spectra of 71 stars in NGC 3680 were obtained at the Cerro Tololo Inter-American Observatory Blanco 4m telescope using the 
Hydra multi-object spectrograph. Three exposures for each of two fiber-position configurations were obtained on 18 and 19 July, 2005.  
One fiber configuration emphasized stars above the turnoff for a total exposure of one hour; the second reached further down the 
main sequence with a total exposure of four hours. Hydra was used in echelle mode; operating at relatively low order (8, in our case),
dedicated filters select and pass the desired spectral region (filter 6 for our purposes). Our spectra were characterized 
by a dispersion of 0.15 \AA\ per pixel in the range 6520 \AA\ to 6810 \AA\  with S/N per pixel that ranged from $\sim$160 to 20. 
Measured widths for calibration lamp lines imply a spectral resolution of $15,000$ in the region near 6700 \AA. 

The numbering system of the proper-motion study by \citet{kp95}(KGP) is one of the larger sets; it will be used in the 
discussions of our own data. All exposures for star KGP 1396 were severely compromised by placement of the spectra on a region of 
the chip affected by a filter defect, so no further discussion of results for this star based on our data will be included. Processing 
of the data and calibration frames was completed at Indiana University and the University of Kansas using standard programs within IRAF.
For clarity, all quoted uncertainties refer to the standard deviation unless explicitly noted otherwise.

\subsection{Radial Velocities}
Radial-velocity and rotational-velocity estimates were derived for 70 stars in our program utilizing the fourier-transform, 
cross-correlation facility, FXCOR, in IRAF. Program stars are compared to stellar templates of similar temperature, in this case 
over the wavelength range from 6575 \AA\ to 6790 \AA\ including H$\alpha$, as well as a narrower region in the vicinity of H$\alpha$ alone.  
A comparison of the results from these two internal measures demonstrates that, if two stars with significant discrepancies are 
excluded from the comparison (one is very cool, the other has very low S/N), the average difference in the radial-velocity determinations, 
in the sense (full spectrum - H$\alpha$), is $-1.27 \pm 2.9$ km/sec. The radial velocity for KGP 802 that appears in Table 1 is based 
on the H$\alpha$ region alone; radial velocities for all other stars are based upon the full spectrum. 

To refer the derived radial velocities to an absolute scale, two options exist. First, two radial-velocity standards were 
observed on each of the nights for the CTIO Hydra runs, HR 9014 and HR 9032. A comparison of our determinations with the 
values from the General Catalog of Radial Velocities \citep{wil53} suggests that our velocities are too large 
by $1.1 \pm 2.1$ km/sec. Second, we can also compare our radial velocities for 18 stars in common with \citet{naa96} 
and 3 stars in common with \citet{merm95}. With known or suspected spectroscopic binaries excluded from the comparison, 
the average velocity difference, in the sense (N96 - this study), is $-0.45 \pm 1.66$ km/sec, indicating that our radial velocities 
are consistent with \citet{naa96} within the errors. 
A similar comparison to three apparently single-star red giants in common with the 
survey of \citet{merm95} produces a velocity difference, in the sense (M95 - this study), of $-0.75 \pm 0.53$ km/sec. We elected not
to correct our radial velocity values based on these fairly noisy comparisons; additional discussion in the next session will suggest that our
radial velocity values are larger than previous results by 0.4 km/sec.

\subsection{Cluster Membership}
Radial-velocity studies by \citet{merm95} and \citet{naa96} established membership credentials for 15 red giants and over 100 red giant 
and dwarf stars, respectively, followed by an analysis of the cluster's evolutionary history \citep{naa97}.  A nearly parallel study 
by \citet{kp97} followed similar lines. The proper-motion study by \citet{kp95} yields membership probabilities for 2711 stars. 
By themselves, the proper-motion measures do not discriminate cleanly between cluster members and field stars for NGC 3680. This point 
is illustrated in Fig. 1, a histogram of the number of stars as a function of proper-motion probability for the 391 stars described 
by \citet{kp95} as ``the most probable cluster members."  From this figure alone, it's difficult to find a sharp 
distinction between members and non-members, except to conclude that stars with $P(\mu) \leq$ 10 to  20\% are unlikely to be members.  
\citet{kp95}, as well as \citet{naa96,naa97}, combine information from radial velocities and proper motions to establish more 
comprehensive membership criteria. In the discussion of memberships for our candidate stars that follows, 
we will employ the proper-motion based probabilities of \citet{kp95} and radial velocities determined from our spectra.

Fig. 2 shows our radial velocities, along with proper-motion-based membership probabilities for our 70 program stars, 
indicating a cluster mean radial velocity near 2 km/sec and illustrating the eventual criteria for cluster membership.  
Whether or not binaries are included, the median radial velocity for stars with $P(\mu) \geq 20\% $ is 1.6 km/sec.  
The average radial velocity for 24 stars with proper-motion membership probability $P \geq 20\%$ that are not known or suspected 
spectroscopic binaries, is $1.64 \pm 1.22$ km/sec. If giants are excluded, the mean becomes $1.72 \pm 1.22$ km/sec. 
The average radial velocity for the four single-star, red giant members (KGP 1175, 1374, 1461 and 1873) 
is $1.28 \pm 0.56$ km/sec.
As expected from the earlier discussion, our mean radial velocity from single member dwarfs is 0.45 km/sec larger than that of \citet{naa97}, who find a mean of 1.27 km/sec from dwarf members.  Our average based on red giants alone is similarly 0.41 km/sec larger than the value 0.87 km/sec found by  
\citet{merm95} from giant members.  \citet{naa97} comment on the difference between the 
average radial velocity of the main sequence and giant stars, and conclude that it is consistent with an expectation based on 
different surface gravities for the two classes of stars; we also find a difference of $\sim 0.4$ km/sec between 
the dwarfs and giants, but the significance of this difference appears to be only marginal in our case. 
\citet{naa97} estimate the internal velocity dispersion from dwarf stars as 0.65 km/sec; using an estimate of our internal accuracy of $\pm0.9$ km/sec leads to a velocity dispersion of 0.8 km/sec.  

Entries in Table 1 include the more common identifications: KGP, as used by SIMBAD, referring to \citet{kp95}; the well-known 
Eggen numbers \citep{eg69}; WEBDA numbers, which use Eggen numbers for the brighter stars and refer to other surveys for the fainter stars.  
Photometric $V$ magnitudes and $B-V$ colors are taken from the collated and merged values as described by \citet{att04}(ATT).  
Estimates of rotational velocity and radial velocity follow, each with error estimates from FXCOR in IRAF. The final column 
summarizes information about binarity. 
Comments about binary status are as specific as possible, including 
notations ``SB?'' referring to the analysis of radial-velocity errors by \citet{kp97} that suggest binary status and, 
similarly, ``w/sb?'' indicating that our spectra suggest unusually wide and/or doubled lines in the instrumental spectra.
Stars with proper-motion probabilities and radial velocities within the bounds 
illustrated in Fig. 2, between $-1.8$ and 5.1 km/sec, are considered members; 
note that while using a modest radial-velocity range to constrain membership, we have 
extended the proper-motion probability limit to 10$\%$. 

\section{Cluster Properties - NGC 3680}
\subsection{Age and Distance Estimates} 
ATT collated and mapped Str\"omgren $b-y$ colors from four other surveys to their own photometry in NGC 3680, 
then mapped the magnitudes and colors to the broad-band photometric system of \citet{kp97}. These are the photometric values 
we will use in the analyses of our spectroscopic data below.  ATT presented independent determinations of the interdependent 
quantities of foreground reddening and overall metal abundance for NGC 3680, as well as extensive reviews of earlier 
determinations of both; the reader is referred to this discussion for details. In summary, NGC 3680 was found to have 
E$(B-V)=0.058\pm0.003$ (sem), well within the range bounded by the $UBV$-based estimate of 0.04 by \citet{eg69} and the high-end 
determination of 0.10 from DDO photometry of the giants \citep{cl83, taat}. 

Photometric metallicity estimates from the intermediate-band $m_1$ and $hk$ indices led to a value of [Fe/H] = -0.14 $\pm$ 0.03 dex (sem), 
lower than the earlier estimates of +0.09 to +0.11 \citep{atts, naa97, bru99} tied to the photometry of \citet{nis88}. The lower
metallicity removed the apparent discrepancy between the photometric dwarf abundances and those of the giants defined by DDO photometry, 
moderate-dispersion spectroscopy \citep{taat}, and high dispersion spectroscopy (PRP). 
Mindful that spectroscopic analyses are necessarily tied to stellar parameters obtained assuming 
an {\it a priori} metallicity estimate, we adopted an initial value of [Fe/H] $= -0.14$ based upon the photometric 
results of ATT but iteratively settled on a value of [Fe/H] $= -0.08$. The rationale behind this choice is discussed in Sec. 3.4 - 3.6. 

With reddening and abundance estimates in hand, a comparison of the homogeneous photometry to appropriate isochrones constrains 
the cluster age and distance, while providing measures of the individual stellar surface gravities. The CMD based on the composite $BV$ 
photometric values (ATT) for 63 probable members, selected using the joint proper-motion \citep{kp95} and radial-velocity criteria 
discussed in Sec. 2.3 from the stars observed for this investigation and those of PRP, is shown in Fig. 3, with open symbols denoting the 
22 stars for which some binary status is known or suspected. Triangles are stars for which only the Li data of PRP are available. 
While the photometric data set is not substantively different from that used by ATT, the isochrones and interpolation software for 
specific ages and abundances adopted for the present discussion are those of Y$^{2}$ (http://www.astro.yale.edu/demarque/yyiso.html) 
\citep{yi03,dem04}, rather than the updated Padova isochrones (http://stev.oapd.inaf.it/cgi-bin/cmd). The latter models \citep{mar08} 
are substantially the same as the \citet{gir02} models used in ATT and require similar offsets of +0.01 mag and -0.023 mag, respectively, 
to the $M_V$ and $B-V$ color scales to bring a one $M_{\sun}$ star of age 4.6 Gyr \citep{ba95, ho08} into alignment with our consistently 
adopted values of 4.84 and 0.65 for the solar $M_{V}$ and $(B-V)$ color. The comparable shifts for the Y$^{2}$ set are $-0.03$ and $+0.008$, 
respectively. The Padova solar isochrones are defined using the Z=0.019 set as the solar composition, while Y$^{2}$ define solar 
as (Y,Z) = (0.266, 0.0181). The switch in isochrone sets allowed a comparison with earlier work to get an indication of how sensitive 
the conclusions are to the specific models adopted, but was explicitly done because the Y$^{2}$ isochrones generate 
masses for the stars in the Li-dip that are more internally consistent with age, a point we will return to in Sec. 4 and 5. For 
Z=0.0152 ([Fe/H] $= -0.08$), Y$^2$ isochrones with ages of 1.7 Gyr and 1.8 Gyr are displayed in Fig. 3, adjusted for the adopted reddening 
of E$(B-V) = 0.058$ and an apparent distance modulus of $(m-M)$ = 10.3. The age and apparent distance modulus are in good agreement 
with those discussed in ATT (1.85 Gyr, 10.2); the revised Padova isochrones yield 1.85 Gyr and $(m-M)$ = 10.2, while supplying a 
slightly better match to the shape of the turnoff below the hook. 

Note in Fig. 3 that the stars traditionally identified as the clump have the predicted luminosity for 
He-core-burning stars in this phase but overlap the colors of the first-ascent giant branch. Moreover, the luminosity of 
the majority of the stars also is consistent with the location of the red giant bump, where first-ascent giants temporarily 
reverse their evolution up the giant branch when the H-burning 
shell crosses the discontinuity in composition created by the deep limit of the convective atmosphere as the star ascends the giant branch.

The solid agreement between observation and theory in Fig. 3 boosts confidence in the derived stellar parameters needed for 
the spectroscopic analysis that can be deduced differentially from the CMD. As exemplified by the Padova comparison, however, even 
assuming a fixed reddening and metallicity, adoption of a different set of isochrones based upon alternate models could result in 
slightly different age, distance, surface gravity and mass estimates for the individual stars. From the detailed discussion in ATT, 
the spread in $(m-M)$ found in the recent literature for NGC 3680 is a byproduct of the combined variation in the derived [Fe/H], the adopted 
reddening, and the choice of stellar isochrones, as exemplified by the values of $(m-M)$ = 10.45 to 10.65 derived by \citet{naa97} and 
\citet{kp97} adopting [Fe/H] much closer to Hyades metallicity for the cluster, as well as alternate sets of isochrones. 

\subsection{Cluster Metallicity - Li}
SPLOT, part of the one-dimensional spectrum analysis suite in IRAF, was used to measure equivalent widths of the line blend 
incorporating the Li-doublet at 6707.8 \AA\ and the Fe I line at 6707.45 \AA; the EW measures are summarized in Table 2, along 
with the estimated S/N for each spectrum. S/N per pixel estimates were based on the modal statistics of the spectra in a ``relatively" 
line-free region between 6615 \AA\ and 6625 \AA.  Li measurements were not possible in some stars: either the line was too weak to be measured or the spectrum was distorted in a way which made measurement implausible. 
Of the eight stars in the latter category, three have S/N per pixel below 30 (KGP 802, 803 and 1610), while the others (KGP 979, 988, 1347, 1410A and 1506) 
are known or suspected spectroscopic binaries whose multiple components confused the Li-line region beyond reclamation. 

The largest set of Li EW measurements and abundance estimates for NGC 3680 is that of PRP, offering abundances for 26 stars, 
including 2 suspected non-members, and expanding on an earlier discussion of four approximately solar-type stars by \citet{ran00}. 
While the larger sample of PRP will supply the comparison of choice for our measurements, the discrepancy between the 
published equivalent widths from the same authors for the four stars based upon the same spectra highlights the 
challenge of comparing results from different groups using different spectra.  The Li EW estimates for the four cool, main sequence stars 
in \citet{ran00} are 8 $\pm$ 2 m\AA\ smaller than those found in PRP, leading to an average Li abundance that is 0.07 $\pm0.01$ dex higher 
in the more recent study. Although PRP discounted the memberships of KGP 1405 and  KGP 1365, 
we include them as possible members. It also seems probable that PRP followed an incorrect 
identification by \citet{naa96} of their star 3001 as KGP 1353; it appears that this star is in fact KGP 1365, with a high probability 
of membership according to \citet{kp97}. 18 stars from PRP overlap with the present study, but only 13 have measurable 
Li lines in both. For these 13 stars, the equivalent widths in PRP are larger by $4.3 \pm 11.1$ m\AA, essentially midway between the 
values of \citet{ran00} and PRP, suggesting that our measurements are on the same system within the errors.

Following \citet{ascon}, we exploit the estimate for the metal abundance of the cluster stars to predict and remove the 
contribution of the Fe I line to the EW measurement for Li.  The corrected equivalent width for the 6708 \AA\ line, listed in 
Table 2 as EW\arcmin, along with the measured line width, temperature, and signal-to-noise estimates for each star provide input 
to a grid of abundance estimates developed by \citet{stein} from MOOG model atmospheres and employed by 
\citet{ascon}. For three stars outside the temperature range of the model grid, MOOG was used interactively to estimate the 
Li abundance for the adopted values of temperature and surface gravity by comparing synthesized to observed spectra. A measured 
line may produce, at best, an upper limit for the Li abundance if EW\arcmin\ is no larger than three times the estimated error, 
as formulated by \citet{del93}; stars meeting this criterion are listed separately in Table 2.

A key component in abundance estimation is the scheme by which temperatures 
are derived for stars. For our analysis, temperatures for the unevolved stars have been based exclusively upon 
the reddening-corrected $B-V$ colors, transformed to the temperature plane using the polynomial relation of \citet{caj}:
 
T$_{e}$ = 8575 - 5222.7 $(B-V)_{0}$ + 1380.92 $(B-V)_{0}$$^2$ + 701.7 $(B-V)$ ([Fe/H]- [Fe/H]$_{Hyades}$)

\medskip
\noindent

For our analysis, [Fe/H] for the Hyades was set at $+0.15$, while the abundance for NGC 3680 was assumed to be $-0.08$. For 
[Fe/H] = 0.0, at a solar T$_{e}$ of 5770 K, the implied $B-V$ color is 0.63, slightly bluer than our normally adopted value 
of 0.65 for CMD fits to isochrones. If we demand that the temperature scale match the redder color, it is 
equivalent to shifting the scale upward by $\sim$65 K. 

Since the relation above is defined by dwarfs in the temperature range bracketing the Sun, it isn't applicable to the 
giants in NGC 3680. For the evolved stars, we converted $(B-V)_{0}$ to T$_{e}$ using the color-temperature relation of \citet{ram05}, 
an update of the dwarf and giant calibrations of \citet{al96, al99}. For comparison purposes, over the dwarf color range of interest, 
the \citet{ram05} scale predicts temperatures systematically cooler by $\sim$25 K than the predicted values from the relation adopted above. 
Final derived values of T$_{e}$ for each star can be found in the second column of Table 2. The derived Li abundances are presented in Table 2, 
with the stars separated according to whether the star had a measurable Li abundance or an upper limit as defined by the 
criterion described earlier.  

\subsection{Li - Comparison with Previous Work} 
PRP employed $B-V$ photometry \citep{naa96} to estimate temperatures for the dwarfs \citep{al96} and giants \citep{al99}. 
From the 20 dwarfs and 6 giants discussed by PRP, the mean temperature differences, in the sense (PRP - this study), are -83 $\pm$ 63 K  
and +27 $\pm$ 57 K, consistent with PRP dwarf colors that are systematically redder by about 0.03 mag while the giants are bluer by 0.01 mag. 
Note that the comparison can be made for the entire sample because we have $B-V$ colors for every star analyzed by PRP, 
not just the overlap with this work.  The offset flip between dwarfs and giants as well as the scatter are products of the lower precision 
$B-V$ colors adopted by PRP. Our colors are based upon a weighted average of a number 
of photoelectric and CCD studies transformed to a common system, resulting in typical uncertainties in the  photometric colors at the 0.003 
to 0.010 mag level.  

To enable a more realistic assessment of the two abundance scales and provide some means of merging the two, we have adjusted 
the Li abundances of PRP to our $(B-V)$-based temperature scale using the prescription for the temperature dependence as supplied by PRP. 
For seven dwarfs with measurable Li in both studies, the mean residual, in the sense (PRP - this study) is -0.02 $\pm$ 0.14 dex. If the star 
with the largest discrepancy is dropped, the average for 6 stars is +0.01 $\pm$ 0.10 dex. The Li abundances for these 7 dwarfs will be based 
upon the average of the two measures, without adjustment for an offset. For the 4 giants with measured Li, the mean residual 
is -0.11 $\pm$ 0.11 dex. This offset has been applied to the data of PRP and averaged with ours. In the following discussions of NGC 3680 
as a representative of intermediate-age open clusters, the Li abundances of PRP for stars not observed in the current investigation will be used 
to augment the cluster sample, with the offsets based upon the dwarf and giant comparisons applied as needed.

Based upon multiple abundance reductions for our sample under different assumptions 
for the temperature scale, we find that a change in T$_{e}$ of $\pm$60 K will alter A(Li) by $\pm$0.05 dex for the dwarf stars. 
For the giants, a comparable temperature change alters A(Li) by $\pm$0.09 dex.

\subsection{Cluster Metallicity - Fe, Ca, Si, Ni}
The determination of the metal abundance of NGC 3680 began with the adoption of a list of cleanly measurable iron lines used in common 
with other cluster investigations by the authors from the Indiana University group.  
To these eight Fe lines we added six measurable lines of neutral Si, Ca and Ni.
Equivalent widths were measured using the IRAF routine SPLOT, then processed using the force-fit abundance routines in MOOG.  These steps 
were replicated for 15 separate exposures of the solar spectrum obtained from our daytime sky exposures from the 2005 Hydra run, adopting 
an effective temperature of 5770 K, surface gravity of log $g$\ = 4.40 and a microturbulent velocity of 1.14 km/sec for the Sun. Signal-to-noise 
ratios per pixel for each of these solar spectra ranged from 90 to 230.  All resulting stellar abundances were differenced with respect to the generated solar 
abundances on a line-by-line basis. 

MOOG provides abundances for each measured equivalent width consistent with an appropriately selected atmospheric model, so the choices 
of atmospheric parameters are critically important.  The choices for temperature scales based on $(B-V)$ colors have been described earlier; 
gravities were determined by examining isochrones fitted to the observed CMD as shown in Fig. 3. For composite systems, 
this will lead to slightly lower gravities than expected for a single star, but the change is too small to 
have a serious impact on the final abundance estimates. Microturbulent velocities for dwarf stars were computed from temperatures and 
surface gravities using the algorithm of \citet{edv}.

Abundances for the giant stars, for which the \citet{edv} algorithm is not applicable, are considerably more sensitive to the choice of microturbulent velocity.  
As a starting point we considered the formula proposed by \citet{car04}: $v_{t} = 1.5 - 0.13$ log $g$, which yields $v_{t}$ values 
between 1.16 km/sec and 1.24 km/sec for the eight giant stars in NGC 3680. However, based upon the standard technique of minimizing the 
trend between the residuals in [Fe/H] and the EW for the lines, PRP found significantly larger values of $v_{t}$ = 1.70 km/sec for their 
coolest giant and 1.90 km/sec to 1.95 km/sec for the remaining five giants. Attempts to apply the \citet{car04} formula to our spectra 
also failed to achieve appropriate minimization of the slope between the residuals and the EW. A range in $v_{t}$ was tested for 
each star until a satisfactory slope was achieved. The resulting trend of $v_{t}$ with log $g$\ preserved the \citet{car04} 
gravity dependence of $v_{t}$ but implied an intercept larger by 0.35 km/sec, midway between PRP and \citet{car04}.

Results for individual stars are presented in Table 3. For the giant stars, all eight Fe lines and the six additional metal lines 
were measurable in all eight stars.  For the dwarf stars, from four to six iron lines were measurable in the chosen stars along 
with the 6717 \AA\ line of Ca, the 6721 \AA\ Si line, and the three Ni lines. The abundances for the dwarf stars exhibited no dependence on T$_{e}$, though the temperature range is modest. An examination of the abundances 
as a function of signal-to-noise, however, suggests that the most consistent dwarf abundances are obtained from the five dwarfs 
with the highest S/N spectra; these five stars are designated in Table 3 by KGP identification numbers in bold font.  
For reference, we note here that we have conformed with what is presumably a standard practice when combining separate measures of [A/H], namely
constructing a simple mean of [A/H] values.  It was anticipated that following this practice would facilitate comparison with other published work, but it cannot be construed as a {\it correct} way to combine logarithmic values.  The preferred approach (taking the logarithm of averaged numerical abundances) yields abundances that are typically larger by amounts that reflect the scatter among individual values.  

\subsection{Metallicity - Giants}

The mean [Fe/H] for 8 giants in Table 3 is -0.17 $\pm$ 0.07, with comparable values for Si of +0.05 $\pm$ 0.12, based 
on one line, and -0.17 $\pm$ 0.08 for the three lines of Ni. A weighted average for [Fe/H] gives 
-0.18 $\pm$ 0.07. While the balance of the measures clearly favors a subsolar abundance for NGC 3680, it leads to the obvious 
question of the zero-point of the metallicity scale since the solar spectra and their analyses are distinctly removed from 
the much cooler and more luminous giants. To test our [Fe/H] estimate, we turn first to the six giants 
analyzed by PRP, five of which have abundances in our survey. PRP obtained 
[Fe/H] = -0.26 $\pm$ 0.11 from 6 giants. The first step in the comparison is to correct for the differences in temperature 
caused by PRP's use of less precise photometry, a smaller reddening, and a different $B-V$ calibration. 
This adjustment produces only a modest decrease in the PRP mean value to [Fe/H] = -0.28, while reducing the scatter to $\pm$ 0.08 dex. 
The truly significant adjustment comes from the difference in the adopted $v_{t}$. For the 5 stars common to the two samples, if we 
had adopted the microturbulence values used by PRP, the mean [Fe/H] would shift by -0.13 $\pm$ 0.05 dex. This is exactly the difference 
and the scatter one gets by taking the average residual for the 5 giants after placing the abundances on the same temperature scale. 
We have shifted PRP's data for the 6 giants by +0.13 dex and averaged our values with theirs for the 5 stars in common to derive 
a mean [Fe/H] = -0.15 $\pm$ 0.08 from 9 giants. The slightly higher value of [Fe/H] comes from the inclusion of KGP 1379 (E34), which 
has the highest abundance of all the giants. We opted not to analyze this 
star since its composite color would likely distort any temperature estimate.  
As one might expect, if this star is dropped, the mean [Fe/H] of the remaining 8 giants becomes -0.17 $\pm$ 0.06.

Two high-resolution spectroscopic investigations of giants in NGC 3680 have been published recently.
\citet{smi08} discussed the giant KGP 1175 (EG13), for which they derive [Fe/H] $= +0.04 \pm 0.11$ (std dev). In spite of their note to the contrary, 
this star is one of the five studied by PRP that overlaps with the current investigation. For their analysis, the stellar parameters 
T$_{e} = 4660$ K, log $g =2.6$, and $v_{t}= 1.3$ km/sec were derived purely from the requirements of internal 
consistency among the line abundances with EW, ionization state, and excitation potential.  Our temperature for this star 
is 4606 K, while PRP get 4668 K. Adjusting to our scale would lower the [Fe/H] to approximately -0.01. Our log $g$\ is 2.56, so no 
correction is required, leaving a difference of 0.12 dex between their estimate and ours for this star. If our derived 
$v_{t}$ of 1.56 km/sec is reduced to 1.30 km/sec, our [Fe/H] is shifted by 0.11 dex, accounting for the remainder of the difference. 

The second paper \citep{san08} details spectroscopy of three single-star giants, 1175, 1374, and 1461, analyzed using two 
different line lists, with the stellar parameters set spectroscopically. The two line lists generate 
different combinations of the stellar parameters for the same star, but the systematic offset between the abundances for the 
three giants is only 0.02 dex, even though the temperature scales differ by 127 K.  The temperature shift appears to compensate in part
for the systematic change in $v_{t}$, which drops by an average of 0.05 km/sec for the lower T$_{e}$ scale. 
As an example of the size of the effects we are discussing, for KGP 1175 on the cooler temperature scale, \citet{san08} 
find (4657 K, 2.68 dex, 1.38 km/sec) and [Fe/H] = -0.01. The expected value with adoption of the $v_{t}$ lowered to 1.30 
km/sec \citep{smi08} is +0.03. 

If we adopt the parameters tied to either temperature scale of \citet{san08} for analyzing our 
spectra for the three giants common to the two studies, [Fe/H] estimates for KGP 1175, 1374, and 1461 are increased 
by +0.12 $\pm$ 0.02 dex for the cooler scale and +0.15 $\pm$ 0.02 dex for the hotter scale. If the offsets are applicable 
to the entire giant sample, our metallicity from the giants becomes [Fe/H] = -0.05 and -0.02, respectively, the same values found by 
\citet{san08}. Note that the very good agreement among the abundance estimates when
the same stellar parameters are applied weakens the argument that the discrepancy among the studies 
is a byproduct of underestimated EW from spectra of lower resolution and S/N compared to those available to \citet{san08}.

The good news, therefore, is that the high-dispersion spectra for giants discussed to date, when processed using the same fundamental 
parameters, will generate Fe abundances that are similar at a level typically $\pm$0.02 dex. The bad news is that for a fixed temperature
scale, the absolute abundance scale is dominated by the adopted microturbulence velocity, $v_{t}$, which leads to a range 
from [Fe/H] $= 0.05$ to $-0.30$ for $v_{t}$ between 1.3 to 1.9 km/sec. 

A potential handle on the location of the zero-point of the metallicity scale comes from the observation by PRP of two stars in the Hyades, 
for which they derive [Fe/H] = +0.06. Since the temperature scale for the Hyades giants as defined by \citet{al99} is virtually identical 
to that of \citet{ram05}, the modification applied to the NGC 3680 abundances due to the altered photometry, reddening, and metallicity doesn't 
apply. Thus, any remaining offset should be a byproduct of the adopted $v_{t}$. Assuming the same formulation applied to 
derive $v_{t}$ for NGC 3680 results in the same systematic offset when applied to the Hyades, if one adopts a metallicity 
for the Hyades near the higher end of the traditional values, [Fe/H] = +0.15, this shifts the PRP corrected abundance for NGC 
3680 from -0.29 to -0.20, still out of range of solar abundance but a respectable match to the value derived in this study. 
We will return to this question again after the discussion of the results for the dwarfs.

\subsection{The Dwarfs}

For the 13 stars at the turnoff with S/N above 70, the average [Fe/H] = -0.04 $\pm$ 0.11, while Si, Ca, and Ni have [A/H] = -0.06 $\pm$ 0.10, 
-0.17 $\pm$ 0.12, and -0.12 $\pm$ 0.11, respectively. As noted earlier, the scatter among the abundances is predictably 
correlated with S/N. If the analysis is limited to the 5 stars in Table 3 with S/N above 110 (boldface ID), the Fe, Si, Ca, 
and Ni abundances become -0.07 $\pm$ 0.02, -0.13 $\pm$ 0.05, -0.22 $\pm$ 0.07, and -0.18 $\pm$ 0.11, respectively. Within the 
uncertainties, the abundances are consistent with a scaled-solar distribution.

For comparison, \citet{pac08} and \citet{san08} have derived abundances for two G-dwarfs in the cluster from the same
spectra, finding [Fe/H] = 0.00 and -0.03 for KGP 1752 and -0.07 and -0.04 for KGP 1050, respectively,
leading to an average of [Fe/H] = -0.04, consistent with either value for our turnoff dwarfs. However, for consistency, 
these estimates deserve closer examination. Unlike the giants, the errors in the dwarf abundances are dominated by the temperature scale and 
are only very weakly dependent upon $v_{t}$. \citet{pac08} use three reddening-corrected color indices coupling $V$ to 2MASS $JHK$ photometry, 
then transferred to T$_{e}$ using the relations of \citet{ram05} and averaged to define a photometric temperature. This value supplies 
the starting point for a grid of stellar atmospheres that are interpolated among the variables until the line abundances show minimal 
trends with EW, excitation potential, or ionization state.

For star KGP 1050, \citet{pac08} determined the photometric temperature to be 6053 K, similar to our $B-V$ based estimate of 6098 K, 
but hotter than the value adopted in PRP by 174 K. However, the spectroscopic temperature used to estimate the abundance was 6210 K, 112 K hotter 
than the scale used in our abundance derivation for the dwarfs. Use of our photometric scale would lower the abundance of KGP 1050 to [Fe/H] = -0.16; 
using the photometric scale of \citet{pac08} would shift the value to -0.20. By contrast, \citet{san08} adopt 6134 K, 36 K hotter than 
our photometric value; use of our scale would reduce their abundance to [Fe/H] = -0.07.

Star KGP 1752 follows a similar pattern in that the spectroscopic temperature (6010 K) \citep{pac08} is 84 K higher than the $VJHK$ average (5926 K), 
but in this case the photometric value is only 10 K hotter than the PRP temperature and 6 K cooler than the estimate in \citet{san08}. 
We have no direct means to tie the star reliably into our temperature scale because its composite $B-V$ color has a dispersion above 0.05 mag, 
one of the largest among the brighter stars in the photometric catalog. If we assume that the $VHK$ value is comparable to our system, 
the abundance for this star would decline to [Fe/H] = -0.06 for \citet{pac08} and remain unchanged for \citet{san08}.   

However, there is a more important reason why the results for this star remain questionable and why it wasn't included in our investigation. 
The star has a high proper-motion membership probability of 58$\%$ and its radial velocity is consistent with being a single-star member of 
the cluster. Despite this, \citet{naa97} flag the star as a questionable member due to its suspicious position in the CMD. 
The photometric and spectroscopic temperatures for KGP 1050 and 1752 \citep{pac08} imply a temperature difference between these two 
stars of 127 K and 100 K, respectively, while the spectroscopic scale of \citet{san08} leads to a difference of 202 K. A typical 
shift of 110 (200) K requires KGP 1752 to be 0.03 (0.05) mag redder than KGP 1050. On the main sequence, being cooler by this amount should 
place KGP 1752 almost 0.2 (0.3) mag fainter than KGP 1050. Instead, KGP 1752 is 0.3 mag brighter than KGP 1050, placing it 0.5 (0.6) mag above its 
expected position in the CMD for a star of its color. If the star is not a binary (for which there is no spectroscopic evidence), it is either a non-member or its structure does not resemble that of a typical solar-type star. For 
either of the above reasons, any spectroscopic abundance tied to this star remains questionable. 

What is the correct metallicity for NGC 3680? From the data obtained solely from this study, the 8 giants have [Fe/H] = -0.17 $\pm$ 0.07 
while the best data for the 5 high S/N dwarfs imply [Fe/H] = -0.07 $\pm$ 0.02, with a simple average of -0.12. If one derives a weighted
average tied to the inverse of the standard errors of the mean, [Fe/H] = -0.09. When analyzed adopting the same parameters selected in 
this investigation, the 6 red giants of PRP and the one giant of \citet{smi08} have [Fe/H] = -0.15 $\pm$ 0.08 and -0.13 $\pm$ 0.10, 
respectively. As emphasized previously, the problem with the giants is that by an appropriate choice of $v_{t}$, the mean metallicity 
could easily be raised to solar or higher. The one definitive dwarf member analyzed outside this investigation, when placed on our 
temperature scale, has [Fe/H] between -0.07 \citep{san08} and -0.16.

From past work, the exclusive metallicity estimate for the dwarfs for over 15 years was the $uvby$H$\beta$-based value 
of +0.1 of \citet{nis88}, reproduced using the same photometric zero-points by \citet{atts} and \citet{bru99}. More recently, 
the $uvbyCa$H$\beta$ CCD study of ATT identified a possible error in the $m_{1}$ photometry of the earlier work and obtained 
a second metallicity estimate using the independently calibrated $hk$\  index. From 34 stars at the turnoff using both photometric 
indices, $m_{1}$ and $hk$, the cluster metallicity is found to be [Fe/H] = -0.14 $\pm$ 0.03 on
a scale where M67 and the Hyades are found to have [Fe/H] = -0.06 and +0.12, respectively \citep{ntc}. For the giants, \citet{taat} 
used the combined observations on the DDO system with the moderate dispersion spectra of \citet{fj93}, transformed to 
a common metallicity scale, to obtain [Fe/H] = -0.10 $\pm$ 0.06 from 12 measurements. On the same metallicity scale tied to DDO 
photometry and moderate-dispersion spectra, 36 measurements for giants in M67 produce [Fe/H] = 0.00 $\pm$ 0.09. \citet{taat} also 
provide a possible avenue for constraining the zero-point scatter among the various high dispersion abundances for the cluster 
dwarfs and giants. \citet{san08} investigate the relative abundances between dwarfs and giants in 13 open clusters, finding a small
effect with metallicity that is readily corrected. Of the 13 clusters, 12 are included in the catalog of \citet{taat}. For these 12
clusters, the average difference in [Fe/H], in the sense (santos - taat), is 0.033 $\pm$ 0.050 dex. The expected scatter from the 
quoted errors for the individual abundance estimates is $\pm$0.046 dex. If the residuals are weighted by the inverse errors, the 
mean offset rises to 0.042 dex. With [Fe/H] = -0.04 from 3 giants in NGC 3680, the result of \citet{san08} becomes [Fe/H] = -0.08 
on the system of \citet{taat} where, again, the NGC 3680 giants were found to have [Fe/H] = -0.10 $\pm$ 0.02 (sem) from 12 
spectroscopic and DDO measurements. 

An average of the various abundance estimates without regard to zero-point and weighted by the inverse of the errors is dominated 
by the dwarfs: [Fe/H] = -0.08 $\pm$ 0.02 (sem). Adoption of values between [Fe/H] = -0.12 and -0.04 have little impact on our 
conclusions regarding the Li abundances. A change in [Fe/H] of $\pm$0.04 will have a small effect on the ages ($\mp$0.1 Gyr), 
distance moduli ($\pm$0.05 mag), and Li-dip masses ($\pm$0.03 $M_{\sun}$).

\section{The Li-Dip: Cluster Data}
\subsection{NGC 3680}
With the establishment of a plausible set of cluster properties for NGC 3680, we can now attempt to place the cluster's Li-dip in the context 
of the physical parameters that define the dip and its evolution over time. The key papers that have dealt with this issue previously 
in a comprehensive manner are \citet{sb95} using open clusters and \citet{che01} evaluating field dwarfs. In a discussion of the pattern 
of A(Li) in several open clusters, \citet{sb95} concludes that the masses of the stars in the dip depend on metallicity, but the ZAMS 
surface temperatures of the stars within the dip do not. The temperature range of the dip appears to depend less on metallicity than 
on age, based on a comparison of Praesepe and the Hyades, two clusters of similar age but supposedly different metallicity, a point 
we will return to below. \citet{che01} similarly find that the location of the dip is metallicity-dependent and derive a relation for 
the mass of the central dip stars as a function of [Fe/H], with solar metallicity dip stars having masses near 1.4 $M_{\sun}$, 
declining to 1.1 $M_{\sun}$ at [Fe/H] = -1.0.

The canonical illustration of the main sequence Li-dip is a plot of Li abundances versus main-sequence 
surface temperature or color. For clusters as old as NGC 3680, beyond illustrating the difference in Li between the main sequence 
and the evolved giants, this representation provides little insight since evolution has shifted the Li-dip into the vertical 
strip at the CMD turnoff \citep{dan94, sb95}. The more functional approach is to use $V$ as the discriminator of mass, 
keeping in mind the potential confusion caused by binary systems. The relevant diagram using our merged and homogeneous sample for
NGC 3680 is shown in Fig. 4. 
Filled circles and triangles represent stars with detectable Li and upper limits, 
respectively, for stars in our sample, averaged with PRP where available. Open circles and triangles show stars with detectable Li and upper limits, 
respectively, from the sample of PRP alone. 
With the sample boosted from 18 to 40 
stars, the features defining the Li distribution among the turnoff stars in NGC 3680 are enhanced and expanded. 
Keeping in mind that binarity can shift the position of a star in the diagram by, at most, 0.75 mag to the right and the
predominance of upper limits for Li estimates within the dip region, the location of the center of the dip is reasonably 
defined as $V$ = 13.45 $\pm$ 0.15. This approximate location is obtained by giving significant weight to the distribution
of measured Li abundances that comprise the wings of the dip rather than the upper limits that dominate the core of the
dip. The midway point between the wings is between $V$ 13.5 and 13.6. Taking into consideration the fact that the
drop in Li is normally somewhat steeper on the high-mass side, this was adjusted to a central dip at 13.45, though clearly
values as high as 13.6 are plausible, as defined by the error estimate. A more exact definition of the location of
the dip center would require an increase in cluster members (not possible) or the superposition of data from clusters 
of comparable age and composition, the focus of Section 5.  

From the minimum in the Li-dip, the Li abundances for higher mass stars rise steeply to an upper bound of A(Li) = 3.35 by $V$ = 12.85, maintaining this high value to at least $V$ = 12.4. There is one star (KGP 1035) at the 
Li upper bound at $V$ = 11.9; it is classed by \citet{naa97} as an SB2 and its position in the CMD places it in an evolutionary 
phase that is so rapid that it is highly likely this star has been shifted to smaller $V$ from a true position between 
$V$ = 12.4 and 12.8 by its composite nature. By contrast, the one star brighter than the red hook in the turnoff which shows 
no evidence of binarity (KGP 1522) has A(Li) = 2.80, almost a factor of 4 below the upper bound. This is consistent with a significant 
Li reduction as stars evolve through the hydrogen-exhaustion-phase toward the red giant branch, where the highest Li abundances 
are reduced by over two orders of magnitude relative to the upper bound \citep{pas04}. When 
contrasting observation and theory for clusters of this age, it's important to remember that the exact shape of the turnoff hook depends very sensitively on the treatment of convective core overshoot in the models. 

On the lower mass side of the dip, the Li abundance rises almost as rapidly, reaching a maximum at A(Li) = 2.75 $\pm$ 0.15 
near $V$ = 14.2 $\pm$ 0.1, with seemingly little scatter. Toward fainter $V$, the Li abundance declines by an order of magnitude, reaching a value 
near A(Li) = 1.8 by $V$ $\sim$ 15.2.

For purposes of constraining the dip and its evolution, the $V$ dependence of the 1.8 Gyr isochrone in Fig. 3 has been converted 
to a mass scale, as illustrated along the top of the plot in Fig. 4. Based on the $V$-mass relation, the ZAMS mass of the stars that 
occupy the center of the Li-dip is 1.35 $\pm$ 0.03 $M_{\sun}$. The current edges of the dip range between 1.51 $\pm$ 0.03 $M_{\sun}$ 
at the high end to 1.19 $\pm$ 0.02 $M_{\sun}$ at the low end. If we were to transform this mass range to a main sequence 
at [Fe/H] = -0.08 at the age of the Hyades (0.7 Gyr), the colors coupled to these mass points would be $B-V$ = 0.40, 
0.32, and 0.50, respectively.

\subsection{NGC 752}
The first and most straightforward comparison for NGC 3680 is with the very similar open cluster, NGC 752. Building upon the extensive discussion 
in ATT, the key issue in this, as in any, comparison is {\it how} similar. There is little disagreement over the relative reddening of the 
two clusters. Typical estimates for NGC 752 range between E$(B-V)$ = 0.02 and 0.04, while NGC 3680 currently is bracketed by 0.05 to 0.07. 
We will assume E$(B-V)$ for NGC 752 = 0.035, consistent with our earlier work. The more contentious issue is the metallicity. An early 
evaluation of the state of the metallicity measures for the cluster including standardization and reanalysis of all available 
photometric data was undertaken by \citet{tw83} and updated and expanded to include spectroscopic data by \citet{dan94}, leading 
to a composite [Fe/H] of -0.15, with significant weight given to 
the spectroscopic results. Photometric zero-point issues aside, this conclusion would imply an abundance comparable to NGC 3680. Due to 
improved calibrations, better zero-point photometric checks, and expanded spectroscopic data, the earlier results have 
gradually been superceded, with a modest change in the answer. 
Using the combined homogeneous results from DDO photometry 
and moderate-dispersion spectra of the giants \citep{taat} in the manner outlined in Sec. 3.6 for NGC 3680, the metallicity of NGC 752 is found to be [Fe/H] = -0.09 $\pm$ 0.07 from 16 measures. It should be emphasized 
in this particular case that, unlike NGC 3680, the DDO result (-0.16) differs from that of the spectroscopy (-0.05) by more than 0.1 dex. A 
significant improvement occurs in the results from $uvby$H$\beta$ photometry, where the $m_{1}$ data of \citet{tw83} exhibited a systematic offset
compared to that of \citet{cb70} and \citet{nis88}, thereby generating an abundance below [Fe/H] = -0.2. With the resolution of the 
zero-point question \citep{jt95}, the mean [Fe/H] of NGC 752 on the $uvby$H$\beta$ system is well established at -0.08 $\pm$ 0.11 from 31 stars, 
on a scale where NGC 3680 has [Fe/H] = -0.17 from $m_{1}$ data alone and M67 and the Hyades are at -0.06 and +0.12, respectively.
Recently, \citet{att06} observed NGC 752 for the first time on the extended $uvbyCa$ system and, from 10 single-star members at the 
cluster turnoff, find [Fe/H] = -0.06 $\pm$ 0.09 on a scale where the Hyades has [Fe/H] = +0.12.

Despite its proximity to the Sun, there have been surprisingly few spectroscopic studies of stars within NGC 752. At high dispersion, 
\citet{ht92} find [Fe/H] = -0.09 $\pm$ 0.05 from 8 turnoff stars on a scale where similar analysis in M67 generates [Fe/H] = -0.04 $\pm$ 0.12 
for 5 dwarfs. The high-dispersion data of \citet{sr752} for three stars imply [Fe/H] = +0.01 $\pm$ 0.04 from 7 dwarfs, 
on a scale where the Hyades has [Fe/H] = +0.14 $\pm$ 0.06, based on 3 stars. A similar analysis from the same group \citep{ran06} 
leads to [Fe/H] = +0.03 $\pm$ 0.03 from 10 stars in M67 on a scale where one Hyades star gives [Fe/H] = +0.13.  While \citet{sr752} dismiss 
all previous metallicity estimates as either irrelevant or unreliable, a careful analysis confirms that they are remarkably consistent, with 
most comparisons of the absolute scale distorted by differences among the metallicity zero-points. A classic example is presented by the
moderate-dispersion work of \citet{fj93} and \citet{fri02} where cluster abundances are derived with a zero-point for the metallicity scale 
defined so that M67 has [Fe/H] = -0.09 in the original analysis and is fixed at [Fe/H] = -0.15 in the latter
paper. Any comparison of the results from these papers to a spectroscopic scale where M67 is observed to have [Fe/H] = +0.03 will lead automatically to a spurious discrepancy if no zero-point correction is applied. 

From the discussion above, the qualitative pattern that emerges is that NGC 752 is more metal-rich than NGC 3680, but more metal-poor than M67. 
A weighted average of the above techniques leads to [Fe/H] = -0.05 $\pm$ 0.04 on a scale where M67 has [Fe/H] = 0.00 and the Hyades 
is at +0.12.

With the reddening and metallicity differential set, we can now make use of the results of ATT to place the clusters 
on the same apparent magnitude scale. By direct superposition of the cluster CMD's, ATT found that the apparent modulus of NGC 752 
had to be 1.90 mag smaller than that of NGC 3680, assuming they had the same metallicity. With a higher [Fe/H], the main sequence of 
NGC 752 should be brighter than that of NGC 3680 by approximately 0.03 mag, reducing the offset to 1.87 mag. 
With the derived $(m-M)$ for NGC 3680 set at 10.30, this implies
$(m-M)$ = 8.43, a boost from the value of 8.30 found in ATT. Using an appropriately interpolated set of Y$^2$ isochrones for the slightly higher 
metallicity adopted for NGC 752, the age of the cluster becomes 1.45 $\pm$ 0.1 Gyr, as shown in Fig. 5, a slight reduction from the value of 
1.55 Gyr derived in ATT. Note that the typical position of the giants is blueward of the first-ascent relation by 0.07 mag.

The combined Li data for NGC 3680 (black symbols) and NGC 752 (red symbols) are shown in Fig. 6, with the data for NGC 752 shifted to 
the distance of NGC 3680. The data for NGC 752 are compiled from \citet{hp86}, \citet{ph88} and \citet{sr752}.
Because each of these NGC 752 studies listed equivalent widths for the Li line, we were able to analyze the data in a manner 
consistent with our treatment of our own measures of Li in NGC 3680, including use of the temperature-color calibration described above.  
A metal abundance [Fe/H] = -0.05 was assumed for this purpose, with a reddening value of E$(B-V) = 0.035$ applied 
to $(B-V)$ colors compiled by \citet{dan94}. For two of the three stars for which equivalent width measurements by \citet{sr752} 
replicated measures by \citet{hp86} and \citet{ph88}, the resulting Li abundance estimates were averaged.  For one of the 
stars, P859/H207, the published equivalent widths are widely discrepant. The measured equivalent width published by \cite{sr752} was used 
to compute the Li abundance for this star.

From Fig. 6, the morphological agreement between the two clusters is impressive. There is a clear gradient in the upper bound 
relation as one moves from the brighter stars to the fainter main sequence. The Li abundance pattern near the solar system Li value
is, within the errors, the same for both clusters among the higher mass stars above the gap. The sample of stars in NGC 752 
with measured Li in this CMD region is frustratingly small; we would predict that a larger sample of observations of stars at the 
turnoff for $V$ brighter than 11.4 in NGC 752 would produce a pattern identical to that in NGC 3680 unless parameters other than age and metallicity are at work.

Moving down the turnoff, the Li-dip is defined at the high-mass edge by a steep drop at $V$ = 13.2, in contrast
with the low-mass edge which shows a gradual rise from $V$ = 13.6 to A(Li) = 2.8 at $V$ = 14.25. 
Toward lower masses, the small sample in NGC 3680 meshes perfectly with the decline observed in NGC 752. One potential point of concern could be
the fact that the Li-dips superpose so smoothly even though NGC 3680 is 0.3 Gyr older than NGC 752 on the current scale. However, the mass of the
stars within the Li-dip is low enough that between 1.45 Gyr and 1.75 Gyr the position of the dip shifts to brighter $M_{V}$ by less than 0.05 mag. 
Taking into account the slightly higher metallicity of NGC 752 relative to NGC 3680, the mass of the dip
stars in NGC 752 should be slightly higher, so the stars should be slightly more evolved at the same
age compared to a cluster with lower [Fe/H]. Thus, we expect that the metallicity effect decreases the impact of the age difference to a
level which is below detectability for the current sample.

\subsection{IC 4651}
As noted in the Introduction, any discussion of NGC 3680 or NGC 752 invariably leads back to the third member of the trio, IC 4651. 
While a common approach is to superpose the Li data of IC 4651 upon that of NGC 3680 and NGC 752, the higher metallicity of this cluster provides 
an equally important handle on the evolution of the Li-dip and the reliability of predictions based upon the adopted 
theoretical isochrones. Based on the masses and projected evolution of stars in the Hyades Li-dip and turnoff, are the observations of the Li-dip stars in the $\sim$1 Gyr older but chemically similar IC 4651 consistent? 

The preliminary step is to place IC 4651 on the same parametric scale as NGC 3680 and NGC 752. Unlike NGC 3680, there has been remarkable 
consistency in the estimation of the cluster parameters from both photometry and spectroscopy. On the photometric side, independent investigations of the cluster on the $uvby$H$\beta$ system, \citet{att87, att00} and \citet{nis88, mei00, mei02}, have exhibited
none of the zero-point issues identified for NGC 3680 and have always indicated that the cluster is similar in metallicity to the Hyades. 
Equally consistent are the reddening estimates tied to the H$\beta$ photometry of \citet{att87, att00} and \citet{nis88}. When analyzed 
using the same intrinsic color relations, the two data sets produce E$(b-y)$ = 0.071 and 0.076, respectively, equivalent to 
E$(B-V)$ = 0.10 and 0.11. With these reddening values, \citet{att00} and \citet{mei02} find [Fe/H] = +0.12 and and +0.13, respectively, 
the latter after adjusting the $m_{1}$ values of \citet{nis88} downward by 0.005 mag to coincide with the 
their CCD survey. Likewise, DDO data for 11 giants leads to E$(B-V)$ = 0.11, and a corresponding [Fe/H] = +0.10 $\pm$ 0.05 \citep{taat}.

On the spectroscopic side, \citet{car04} analyzed 4 giants to derive [Fe/H] = +0.11 $\pm$ 0.01, assuming E$(B-V)$ = 0.083; adoption of a higher 
reddening raises this estimate. \citet{pas04} discuss high-dispersion spectroscopic analysis of over 20 stars ranging from giants to G dwarfs 
and obtain [Fe/H] = +0.10 $\pm$ 0.03 with an adopted E$(B-V)$ between 0.12 and 0.13. Finally, from 3 giants and 5 dwarfs, \citet{san08} derive 
[Fe/H] = +0.15 $\pm$ 0.01 and +0.15 $\pm$ 0.02, respectively. 

For the apparent distance modulus, the consistency among recent determinations tied to main sequence fitting has been equally 
impressive, with one notable exception which will be discussed below. \citet{taat} find $(m-M)$ = 10.25 for [Fe/H] = +0.1 and E$(B-V)$ = 0.11 
and ATT derive $(m-M)$ = 10.20 with reddening reduced to 0.10. Given the similarity of the cluster metallicity to the Hyades, one can avoid 
the usual issues with isochrones by doing a direct comparison to the Hyades. Using the $Vby$ data, \citet{att00} find $(m-M)$ = 10.30 
for E$(b-y)$ = 0.071 (E$(B-V)$ = 0.10) and from the $Vvy$ diagram to much fainter magnitudes, \citet{mei00} finds $(m-M)$ = 10.36 
for E$(b-y)$ = 0.076 (E$(B-V)$ = 0.11). 

While the cluster parameters are exceptionally consistent internally, one minor discordant note has been raised by recent high-dispersion 
spectroscopic work on the cluster. \citet{pas04}, \citet{bi07} and \citet{pac08} have used temperatures based upon consistency constraints on abundances from 
the multiple lines of differing equivalent width, ionization state, and excitation potential to imply that the reddening value of IC 4651 is 
closer to E$(B-V)$ = 0.12 to 0.13. This conclusion is contradicted by a similar approach applied to the red giants by \citet{car04} and
severely weakened by the analysis of \citet{san08}, who generate the same abundances but two distinctly different temperature scales
for the same stars using two different line lists. Ultimately, the differences between the higher reddening value and 
E$(B-V)$ = 0.10 discussed in \citet{att00, att04, mei02} and E$(B-V)$ = 0.11 discussed in \citet{mei00} 
and \citet{taat} are minor and well within the error bars of both approaches. If we adopt E$(B-V)$ = 0.12 for IC 4651, E$(b-y)$ becomes 0.085; 
shifting from the older average of E$(b-y)$ = 0.073 leads to a correction to $m_{1}$ of an additional 0.004 mag, raising [Fe/H] for IC 4651 on 
the Str{\"o}mgren system to [Fe/H] = +0.16, again, not a statistically significant change from earlier results. With 
a reddening of E$(B-V)$ = 0.12, the apparent modulus rises to $(m-M)$ = 10.40 $\pm$ 0.05. Under the assumption that IC 4651 has a metallicity indistinguishable from that of the Hyades and using an appropriately interpolated set of Y$^2$ isochrones ([Fe/H] = 0.15), 
E$(B-V)$ = 0.12 and $(m-M)$ = 10.40, as illustrated in Fig. 7, the age of the cluster becomes 1.50 $\pm$ 0.1 Gyr, a reduction from the 
value of 1.75 Gyr derived in ATT and 1.7 Gyr in \citet{mei02} due to the higher adopted reddening, a slightly higher absolute [Fe/H], and
exact interpolation of the isochrones. Triangles are stars known to be spectroscopic binaries; known non-members have been removed 
from the sample which is a combination of the BV photographic and CCD data in \citet{att88}. It should be noted that the color and 
magnitude offsets applied to the isochrones for NGC 3680 have also been applied to the sets at higher [Fe/H]. A consistency check has 
been made by matching the photometry of the Hyades \citep{jt06} to an isochrone of appropriate age and $(m-M)$ = 3.33, with the 
offsets included; the agreement is excellent. 

The comparison between the data and the isochrones is encouraging, especially at the red hook in the turnoff which is dominated by 
the assumed degree of convective overshoot mixing. Note, however, that the giants all lie blueward of the first-ascent giant branch, 
in sharp contrast with NGC 3680 and as predicted if the giants are dominated by He-core-burning stars. This difference with NGC 3680 
could be removed by altering the reddening of IC 4651 to a value closer to E$(B-V)$ = 0.00, thereby requiring a reduction in the 
apparent modulus by 0.6 mag and significantly boosting the cluster age. As discussed previously, however, reductions in the reddening 
go opposite to the trend from spectroscopy in recent years.  

Before the IC 4651 data are linked to younger clusters of the same metallicity, the particularly anomalous results of \citet{bi07} need 
to be addressed. The focus of their study is the use of line-depth-ratios from high-dispersion spectra of giants to derive temperatures 
which, when combined with absolute magnitude and metallicity, can generate a mass and, ultimately, through isochrone comparisons, 
an age. The problem is that \citet{bi07} quote an age of 1.4 Gyr and an apparent modulus of $(m-M)$ of 9.83 for IC 4651 from a 
comparison to the Padova isochrones. We have reproduced the comparison of Fig. 7 using properly interpolated and zeroed Padova 
isochrones and find excellent agreement with the age and modulus obtained from the $Y^2$ isochrones, as expected from the comparison with ATT.  
The \citet{bi07} age estimate should be adjusted downward to correct for their use of [Fe/H] = 0.10, if their results are to be compared 
with our adopted [Fe/H] = 0.15. Clearly, however, an estimate of $(m-M)$ = 9.8 is implausible given the discussion above and the 
fit in Fig. 7; the minimum  modulus using the lower limit to the reddening, E$(B-V)$ = 0.085, is 10.15, as derived in \citet{att00} and 
adopted by \citet{car04}, and can only increase for higher reddening. \citet{bi07} make no note of 
this apparent anomaly, citing only two dated references \citep{nis88, kf91} that are consistent with their result. 
The primary source of the problem comes from their use of the photometry of \citet{pi98} to compare the cluster CMD 
to the theoretical isochrones. As discussed in \citet{att00}, the $V$ magnitudes in this survey are systematically too bright by 0.20 mag. Even worse, the $V$ zero-points between the two 
CCD fields studied in the cluster differ by 0.09 mag. Because no additional $V-I$ colors are available for comparison, any 
attempt to use their $V-I$ CMD to derive an age by matching to isochrones must be
highly questionable, making the approximate agreement between 1.4 Gyr and Fig. 7 purely fortuitous.

\subsection{Hyades and Praesepe}
As a younger Li standard for comparison purposes, we have combined data for the Hyades and Praesepe. There is no question  
that the Hyades is metal-rich, with [Fe/H] values noted in this paper between +0.12 and +0.17; we have selected +0.15 for the temperature scale. 
What matters, however, is that IC 4651 and the Hyades be the same within $\pm$0.05 dex. The more controversial cluster is Praesepe. 
Following the same sources discussed for the other clusters, DDO photometry of 4 giants produces [Fe/H] = +0.14 $\pm$ 0.07 
and E$(B-V)$ = 0.0 \citep{taat}. $uvby$H$\beta$ photometry of \citet{nis88} implies E$(B-V)$ less than 0.01 and [Fe/H] = +0.10 $\pm$ 0.15 
from 42 dwarfs on a scale where the Hyades has [Fe/H] = +0.12, consistent with the earlier results of \citet{cb69}. \citet{jt95,jt07} 
derive an offset of $\sim$+0.01 mag to the $m_{1}$ index of \citet{cb69}, which would imply a higher [Fe/H]. Washington photometry of 4 
giants in each of the Hyades and Praesepe clusters leads to +0.07 and +0.04, respectively \citep{gcm91}.

Spectroscopically, the high-dispersion standard for some time has been tied to three related papers \citep{bb88, boe89, fr92}. The first 
investigation found [Fe/H] = +0.13 $\pm$ 0.07 (sem) from 5 single dwarfs, on a scale where F dwarfs in the Hyades 
had [Fe/H] = +0.17 $\pm$ 0.06 (sem). The latter two studies generate a combined [Fe/H] = +0.06 $\pm$ 0.06 from 7 dwarfs on a 
scale where the Hyades has [Fe/H] = +0.13; a weighted average combining these two samples raises the Praesepe value 
to +0.09 $\pm$ 0.03 (sem) \citep{an07}. More recently, \citet{an07} used 4 dwarfs to 
derive +0.11 $\pm$ 0.06, on a scale where the Hyades has [Fe/H] = +0.13, while \citet{pac08} obtain +0.27 $\pm$ 0.10 from 7 dwarfs. As a 
final note to the summary from the literature, \citet{tay06} has attempted a rigorous assessment of the reddening of the Hyades, Coma, and 
Praesepe and concludes that while the former two clusters have effectively no reddening, Praesepe has E$(B-V)$ = 0.027. His reevaluation 
of past high-dispersion work in this context implies [Fe/H] = +0.01 $\pm$ 0.04 for the cluster.

With the exceptions of \citet{tay06} and \citet{pac08} at opposite ends of the metallicity limits, there is general agreement that Praesepe 
is metal-rich compared to the Sun and similar to, or slightly lower than, the Hyades. This meshes well with the high-dispersion analysis of a few dozen Praesepe 
dwarfs by the Indiana group which, when processed on the same temperature scale as NGC 3680 with E$(B-V)$ = 0.0, leads to 
[Fe/H] = +0.144 on a scale where the Hyades has [Fe/H] = +0.15.

The results for the combined Li data for the Hyades and Praesepe \citep{stein} are shown as a function of $B-V$ in Fig. 8. Blue circles 
and red squares are the Hyades and Praesepe data with measured abundances, respectively, while the blue and red triangles are upper limits. 
The original EW data for Li in the Hyades are from \citet{tho93, bbr88, bt86, so90}, while the 
Praesepe data are from \citet{bb88, so93, bc98}. The EW data have been converted to A(Li) using the same approach adopted for NGC 3680. 
The $B-V$ colors for Praesepe are from \citet{jo52} and \citet{me67}; both sets of photometry are in excellent agreement with each other. 
For the Hyades, we have adopted the photometry of \citet{jt06} whenever possible. \citet{jt06} identify the need for a zero-point shift
of +0.008 mag for the more commonly used data from \citet{jk55}; for stars for which only the photometry of \citet{jk55} is available, 
this offset has been applied.

Though it is by no means perfect, the agreement between the superposed data is impressive, especially in the region of the Li-dip. 
The one obvious disagreement is in the $B-V$ region between 0.55 and 0.6. Praesepe exhibits a population of stars that has clearly 
depleted the surface Li by a significant amount compared to the well-defined upper bound in this color range. The depletion is 
non-uniform among the Praesepe stars, with one star reaching a factor of 20 below the expected value, and it is not present in the 
less populous sample from the Hyades. In the region of the Li-dip, however, we have drawn by eye a mean relation through the combined sample, 
as illustrated by the solid line. Within the center of the Li-dip, the points defining measured Li abundances have been given 
significantly more weight than the points with only upper limits. Three conclusions emerge from the mean relation:
(a) the $B-V$ center of the Li-dip is the same for both clusters at 0.44 $\pm$ 0.01. For this metallicity and age, the isochrones predict 
that the stars at this point will have a mass of 1.421 $\pm$ 0.015 $M_{\sun}$; (b) the high mass limit of the dip occurs precipitously at 
$B-V$ = 0.405 $\pm$ 0.005, equivalent to 1.494 $\pm$ 0.008 $M_{\sun}$ while the low mass boundary occurs at $B-V$ = 0.517 and a 
mass of 1.28 $M_{\sun}$; and (c) the Li-dip is not symmetric. In particular, there is evidence that in the color range from $B-V$ = 
0.47 to 0.52, Li depletion has occurred at a higher rate than for stars just redward of this region, creating a secondary dip among 
these stars on the main sequence.

If we assume that the Li-dip profile as defined by the Hyades and Praesepe represents that of IC 4651 at the same age, we can now translate 
this relation to the age, distance, and reddening of IC 4651 and see how they they compare. The initial step is the evolution of the
Li-dip to the age of the cluster. This is done by converting the Li-$(B-V)$ relation to a Li-Mass relation, then converting it to
a Li-$M_V$ relation at the age of IC 4651. With the apparent modulus known, this readily translates into a predicted location for the 
Li-dip in the CMD turnoff region. The expectation is that the center of the dip should occur at the appropriate magnitude but, 
at best, the IC 4651 data should lie at or below the mean relation at a given $V$ due to potential depletion in Li between an 
age of 0.7 Gyr and 1.5 Gyr. The Li data for IC 4651 come from three sources, \citet{pas04}, \citet{sb4651} and 
unpublished data from Suchitra Balachandran (private communication). Of these several sources, only \citet{pas04} lists measured equivalent 
widths as well as final abundances and stellar temperatures. We employed $(B-V)$ colors from \citet{att88} to compute temperatures for the 
dwarf stars consistent with our analysis path for NGC 3680 and NGC 752, using a reddening value E$(B-V) = 0.12$ and an assumed metallicity of
[Fe/H] = 0.15. As \citet{pas04} list lithium abundances derived for two separate temperatures, one based on photometric colors, the other 
based on spectroscopic reductions, we were able to adjust the published Li abundances to conform with our temperature 
scale by applying a correction $\Delta {\rm A(Li)} = 7.1 \times 10^{-4} \Delta T$.  Results for stars common to more than one source were 
averaged if both results were derived abundances; if either or both results were upper limits, the more stringent upper limit was adopted.

Initial attempts to superpose the predicted Li-dip distribution upon the observed trend in IC 4651 made use of the Padova isochrones 
and failed in that the predicted location of the Li-dip at 1.5 Gyr was 0.4 mag fainter than observed, implying that the mass center 
of the Li-dip for the Hyades and for IC 4651 were different. Part of the issue may be tied to the need to adjust the Padova isochrones 
to our adopted solar color and luminosity, shifting the color for a star on the main sequence at a given mass. While the combined shift 
was tested for the Hyades and confirmed in that it gave the appropriate distance modulus, it may well be that the relative required 
combined shifts in $B-V$ and $V$ are different at higher [Fe/H] from what they are at solar metallicity, thereby altering the mass-color 
relation. The switch to the Y$^{2}$ isochrones was made, in part, because the offsets applied to the isochrones are smaller and 
conceivably would be less affected by alterations at metallicities different from solar. This has proven to be the case.
Fig. 9 shows the Li distribution for the stars in IC 4651 (open circles are detections while triangles are upper limits) with 
the Hyades/Praesepe Li-dip relation evolved to an age of 1.5 Gyr and shifted by 10.25 mag, identifying the center of 
the Li-dip as $V \sim 13.35$. The mass of the stars at this location is 1.47 $M_{\sun}$, larger than the Hyades-based 
1.42 $M_{\sun}$ and a predicted Li-dip location of $V$ = 13.52. Note that a star with 1.47 $M_{\sun}$ would have $B-V$ = 0.416 
at the metallicity and age of the Hyades, only slightly bluer than our adopted Li-dip center of $B-V$ = 0.44.

Given the small number of stars used to define the IC 4651 profile, the consistency between the shift predicted by the isochrones and the
observations is encouraging. On the brighter, high-mass side of the dip, there appears to be no decline from A(Li) values 
near the solar system value until one reaches the stars evolving through the hydrogen-exhaustion phase and toward the red
giant stage. Within the broad uncertainties, the sharp drop seen on the high-mass side remains unchanged. On the low-mass, 
fainter side, all the cluster stars fall on or below the Li-dip, as expected if some depletion has occurred from the upper
limits for these stars as defined by the Hyades relation. Note also that the secondary dip identified within the
Hyades will be difficult to verify due to the small sample populating the correct magnitude range. 

\section{Lithium in Open Clusters}
\subsection{Li-Dip: Structure and Evolution}
To optimize the insight gained from the clusters, we combine Fig. 7 and Fig. 9 in Fig. 10. Because the metallicities of
NGC 3680 and NGC 752 are so similar and the age difference partially compensates for any evolutionary effect, no
adjustment is made to the relative positions of the points (blue) in Fig. 7. The combined samples for these two
clusters are at an effective true modulus of $(m-M)_o$ = 10.12; if the metallicity is lowered to [Fe/H] = -0.12, the
true modulus is reduced to 10.07. For IC 4651, the true modulus is $(m-M)_o$ = 10.03, identical within the uncertainties
to the distance to NGC 3680. For consistency, however, we have added 0.09 mag to the $V$ magnitudes for the data points in IC 4651 (red) 
and the mean relation (solid line) of Fig. 9 to place them at the same distance as the other clusters in Fig. 10. 

Clearly, the Li-dip profile as defined by the combined Hyades and Praesepe data and evolved to an age of $\sim$1.5 Gyr does an excellent job of mapping
out the region of the observed composite Li-dip. Several key features are  enhanced in the composite sample.
(1) The upper bound at A(Li) = 3.35 for the stars above the Li-dip is well established. One star at the bright edge of the Li-dip 
sits above the upper bound by only 0.1 dex, an insignificant offset given the errors;
(2) Keeping in mind that non-interacting binaries will be shifted 
horizontally to the right and that the sample of stars above the Li-dip is incomplete, the upper bound appears to end abruptly 
at $V$ = 12.35. Stars brighter than this level are leaving the turnoff for the giant branch, either in or beyond the hydrogen-exhaustion 
phase. It should be emphasized, however, that not every star within the main sequence red hook exhibits reduced Li; 
(3) The vertical drop in A(Li) on the high mass side of the dip occurs over a very small range in $V$, consistent with the sharp 
Hyades/Praesepe drop with color. Likewise, the profile on the low-mass side matches the observational upper bound extremely well. 
This appears to indicate that over the $\sim$1 Gyr of evolution between the Hyades and the average age of these clusters, little 
significant alteration has occurred in the Li distribution within the Li-dip, implying that the
physical process producing the dip may have run its course by 0.7 Gyr. It should be remembered, however, that due to the
steepness of the profile and the logarithmic basis of the plot, a drop in Li by a factor of 2 or less would be difficult to
detect; (4) With the exception of a handful of stars near $V$ = 14.7 in NGC 752, the average star fainter than the Li-dip is typically 0.25 dex 
lower in A(Li) than the mean relation for the Hyades, though this differential
would be cut in half if the mean dip relation were shifted to brighter $V$ magnitudes by 0.15 mag. Within the errors, there does appear to be
a modest decline in A(Li) for the cooler dwarfs outside the dip between the ages of the 0.7 and 1.5 Gyr. 

To close our discussion of the Li-dip, we can estimate the mass dependence of the center of the dip on metallicity. Our comparison of the
Hyades/Praesepe dip ($M = 1.42$ $M_{\sun}$) with the evolved location in IC 4651 ($M = 1.47$ $M_{\sun}$) implies that, due to the uncertainties 
in the evolutionary models, independent of the uncertainty in the exact placement of the center of the dip for the older clusters, it is possible 
that the isochrones will predict too high a mass when matched to the CMD position of the clusters near 1.5 Gyr. For NGC 3680 and NGC 752 at an 
approximate [Fe/H] = -0.07, the mass of the Li-dip center is 1.35 $M_{\sun}$. Contrasting this with IC 4651, a linear relation over this 
metallicity range would be $M_{dip}/M_{\sun} = 1.38 (\pm 0.04) + 0.4 (\pm 0.2$)[Fe/H]. For comparison, using field stars over a much 
wider range in [Fe/H] and a different set of stellar models, \citet{che01} found $M_{dip}/M_{\sun} = 1.42 + 0.28$[Fe/H].

What is the physical origin of the Li dip and the regions surrounding it?  Once the Hyades and field dwarf Li data of \citet{bt86,bt86b} firmly established that F stars depleted far more Li than predicted by standard theory (see discussion in \citet{cd00}), a variety of mechanisms beyond the SSET were proposed.  These loosely fall into three categories: mass loss, diffusion, and slow mixing due to rotation or waves.  
In the mass loss scenario, the small outer mass fraction where Li is preserved is simply lost \citep{ssd}.  
In the original diffusion scenario for the Hyades \citep{m86}, diffusion is unimportant for surface Li depletion in G dwarfs, but as one goes to the hotter early F dwarfs, the increasingly shallow 
 SCZ allows increasingly efficient draining of the Li out of the base of the SCZ via gravitational settling and thermal diffusion.  This creates the red side of the gap.  In the middle of the Li gap, Li retains an electron and with its now much larger cross-section, Li becomes radiatively accelerated upward, creating the hot side of the gap.  However, this can cause large Li overabundances in stars hotter than $B-V=$0.40 (a "Li peak"), which was not observed, so a finely-tuned mass-loss rate was included in such stars to balance the diffusion.\footnote[2]{Li peak stars may exist afterall: \citet{caj} discovered a star in the Hyades-aged cluster NGC 6633 near 7100K with A(Li) = 4.3, at least an order of magnitude above the assumed initial abundance of the cluster.  This is consistent with the diffusion predictions of \citet{pm93}, although other explanations are also possible \citep{lg,ajs}}.  The more refined diffusion models of \citet{pm93} predicted a much narrower Li dip ($\sim $100K in T$_{e}$), which could no longer explain the full width of the observed Li dip ($\sim$ 300K for the primary dip and 500K or more when including the secondary dip), and even less so the more widespread Li depletion hotter and cooler than the dip.  

Mechanisms that can drive slow mixing include, a) waves caused by the SCZ \citep{gls,tc}, and b) rotation-related effects such as meridional circulation or the onset of angular momentum transport induced primarily by secular shear (\citet{pdk,cdp,dp}, collectively referred to as the Yale models).  \citet{cm} proposed that meridional circulation combined with diffusion could explain the Li dip, though later work suggested that the efficiency of meridional circulation must be greatly suppressed \citep{cmp,pdk}.  Another candidate process, turbulent transport, overdepletes stars beyond the hot side of the Li dip when diffusion is allowed for \citep{cm}.  In the Yale models, surface angular momentum loss acting primarily during the early main sequence of G dwarfs exacerbates differential rotation with depth, which sets off a secular shear instability, causing mixing.  Different initial rotation rates cause the observed dispersions in Li abundances.  Increasing initial angular momentum with mass and corresponding greater mixing results in the cool side of the dip.  But stars progressively hotter than the break in the Kraft curve (which coincides with the middle of the Li dip) lose progressively less angular momentum, thus producing the hot side of the Li dip.

These various classes of models meet with varying degrees of success in reproducing the Hyades Li dip, and distinguishing between them using the Hyades dip alone can be challenging.  Fortunately, there exist additional powerful discriminators, that include, a) the timing of the formation of the Li dip and the evolution of its Li-mass morphology, b) the Li-Be and Be-B depletion correlations, c) short-period-tidally-locked binaries, and d) subgiants evolving out of the Li dip, whose deepening convection zones reveal the shape and depth of the Li preservation region of their main sequence progenitors.  

The early onset of Li depletion by factors of 4 to 6 and over a wide T$_{e}$ range (6000 to 6700K, at minimum) observed in the 150-Myr-old cluster M35 \citep{ascon} favors the Yale rotational models as the dominant effect at early ages and argues against mass loss or diffusion which require longer time scales.  The striking Li-Be depletion correlation discovered in field dwarfs on the cool side of the Li dip \citep{dbs,bdsk} and in open clusters \citep{bakd} argues strongly in favor of slow mixing induced by rotation, as in the Yale and other models that produce the specifically observed Li/Be ratio (e.g.\citet{cvm}), and argues strongly against diffusion (where Li and Be are depleted from the surface at a similar rate) and mass loss (where all the Li must disappear before Be depletion can be observed).  The Li-Be correlation can even discriminate among models with slow mixing; for example, the wave-driven mixing in the \citet{gls} models does not penetrate very deeply, and as a result the predicted Li/Be ratio is too large.  The wave-driven mixing of \citet{tc} is more promising, especially in its association of angular momentum transport by waves, and in its potential to explain the spread of Li abundances as a function of T$_{e}$ for stars cooler than the Li dip, but it remains to be seen whether more quantitative models can reproduce constraints other than those from Li in the Hyades alone (such as, but not limited to, the constraints described here).  Since Be is preserved to roughly twice the depth as Li with B preserved to a depth twice as large again, the observed Be-B depletion correlation further illustrates the depth to which mixing occurs, and further defines its relative efficiency with depth \citep{cdp}. Combining the Yale rotational models with \citet{zb}'s tidal circularization theory results in the prediction that tidally locked binaries with sufficiently short period should exhibit higher Li abundances than single stars because they exchanged angular momentum during the early pre-main sequence when their interiors were not yet hot enough to destory Li.  Thus, they do not exhibit the subsequent mixing and Li depletion induced by angular momentum loss that normal, single stars do.  (See also \citet{so90}.) This prediction is supported by the observation of high-Li short period binaries above the late-F/early G Li plateau and elsewhere in the Hyades \citep{tho93}, in M67 \citep{del94}, and in a variety of other contexts \citep{rd95}.  Li in subgiants in the solar-aged M67 cluster reveal that the Li preservation region during those stars' MS evolution was smaller that that predicted by standard or diffusion models, consistent with that predicted by the Yale rotational models, and larger than that predicted by mass loss.  These considerably varied diagnostics all consistently favor slow mixing (probably induced by rotation) as the primary cause of Li depletion in F dwarfs, and consistently argue against mass loss or diffusion.  

     The present study adds additional interesting constraints that successful mechanisms must meet.  The possible delineation of the Li dip into a deep primary and shallower secondary Li dip in the Hyades+Praesepe data, which is also consistent with the Li data in the older clusters NGC3680, NGC 752, and IC4651, suggests that Li depletion occurs fairly early (before 200Myr), over a wide range of T$_{e}$ (at least 6000-6700K) and is caused primarily by rotationally-induced mixing, but a more severe Li depletion continues to develop in a more restricted T$_{e}$ range near 6700K later but by the age of the Hyades.  This further depletion could be caused by or be aided by diffusion.  Although two stars with very low Li in the primary Hyades Li dip are also known to have a significant Be depletion \citep{boe02}, consistent with rotational mixing, the Li data are only upper limits, and thus the Li/Be is not known in these stars, and could allow for the possibility of late Li diffusion, especially at a T$_{e}$ where the picture gets further complicated by the possibility of upward radiative acceleration of Be \citep{pm93}, even as Li continues to diffuse downward.  Additional Li data, preferably detections, in the deep parts of the primary Li dip, ideally along with Be data, could help cast light on whether diffusion ever contributes to the formation of the Li dip.  In NGC 3680 stars hotter than the Li dip, we stress that we have found a Li spread of at least 1 dex in magnitude.  That is, these stars do {\it not} exhibit a constant, presumed undisturbed initial cluster Li abundance.  (However, a larger fraction of IC 4651 stars, indeed most of them, are found near the Li upper bound; only three stars have been observed in NGC 752, but they also lie close to the upper bound.)  Such a spread in NGC 3680 is not altogether surprising, as the SCZs are {\it extremely} shallow in these stars, and many complex competing factors can affect the surface Li abundances, including rotationally-induce mixing of varying types, diffusion, magnetic fields, and even just tiny rates of mass loss.

\subsection{Li in the Giants}
While the focus of our study has been the turnoff dwarfs, the added data on the giants in NGC 3680, as well as additional data 
in IC 4651 \citep{pas04}, promise potential insight into post-main-sequence evolution of Li as discussed by ATT and \citet{pas04}. 
For NGC 3680, the sample of PRP has been increased from six to nine giants; all three members of the expanded sample 
are probable binaries. As emphasized in ATT, the surprising result from the earlier work of PRP was the discovery that while 
the Li abundances among the giants of NGC 3680 were reduced relative to the turnoff stars, of the six giants studied, four had 
measurable abundances rather than upper limits, and three of the four measurable values sat near A(Li) = 1.0. This led to a bimodal 
distribution in A(Li). The combined sample confirms and expands the result. Of the three new giants, 2 have A(Li) = 0.98 and 1.09, while 
the third has 0.50. For the 9 giants, five have an average A(Li) of 1.08 $\pm$ 0.08, while the remaining 4 have A(Li) = 0.45 $\pm$ 0.07 or 
an upper limit measurement. It should be noted that the true limit for KGP 1379 is undoubtedly lower since the unusual blue color of this 
star implies a likely main sequence turnoff companion. While the Li line strength would be weakly impacted by the companion, the effect 
of the bluer color is to raise the effective temperature, producing a significant overestimate for Li.

How does NGC 3680 compare to NGC 752? First, of the three clusters, NGC 3680 is clearly the oldest, while NGC 752 and IC 4651 
are virtually identical in age if one accepts the higher reddening value for the latter cluster. As discussed in ATT, NGC 752 and 
NGC 3680 have distinctly different A(Li) distributions among the giants. For NGC 752, of 11 highly probable members on the giant branch, 
only 2 have detectable Li abundances, both above A(Li) = 1. The remaining nine stars have only upper limits of 0.5 or less. 

Beyond the age difference, the second way in which NGC 752 and NGC 3680 distinctly differ is in the CMD distribution of the giants. A comparison
of Fig. 3 with Fig. 5 shows that the giants in NGC 3680 sit squarely on the first-ascent giant branch in the luminosity range associated with the
red giant bump. By contrast, while the giants in NGC 752 occupy a comparable range in luminosity, they sit systematically blueward of the 
first-ascent
giant branch and 0.07 mag bluer in $B-V$ than the typical star in NGC 3680, occupying the CMD area commonly associated with the red giant clump 
and He-core-burning stars. Finally, the two stars in NGC 752 with measurable Li do not sit among the stars populating the clump, but lie along the red
edge of the CMD distribution (see Fig. 8 of ATT). A qualitative explanation comes from the observation that some stars evolving 
through the red hook
at the turnoff exhibit reduced Li from the limiting value of 3.35. Note that of all the stars with measured Li in the three clusters, 
only one star in
IC 4651 is situated in the subgiant region of the CMD between the turnoff and the vertical first-ascent giant branch. It has A(Li) = 1.5. 
All other single stars with reduced Li populate the red hook. The implication is that some form of unexpected mixing, possibly due to 
enhanced rotation \citep{pas04}, is destroying the atmospheric Li as the star evolves away from the main sequence and up the giant branch. 
Thus, first-ascent red giants may still retain a detectable level of Li near A(Li) = 1 as they pass through the red giant bump phase and/or bring 
new Li to the surface as a byproduct of the interior evolution and mixing that creates the red giant bump. Evolution to the He-core-burning 
phase reduces A(Li) of 0.5 or lower, unless an alternative source of Li production is found. When coupled with the timescales
in the different evolutionary phases, giant branch abundances should show a natural dichotomy. One group of predominantly first-ascent giants 
will have measurable Li near A(Li) = 1, while the more evolved stars will have A(Li) = 0.5 or less. NGC 752 is dominated by the latter group, 
NGC 3680 by the former.

Given the pattern above, it is straightforward to predict where IC 4651 should fall in the Li-dichotomy. Its age is effectively identical 
to NGC 752 but the mass of a typical star on the giant branch (1.93 $M_{\sun}$) is higher than that for NGC 752 (1.86 $M_{\sun}$) or NGC 3680
(1.75 $M_{\sun}$). The stars on the giant branch, especially with the increased cluster reddening, sit squarely blueward of the first-ascent
giant branch by more than 0.1 mag in $B-V$ (Fig. 7). The Li distribution among the giants should resemble that of NGC 752. As discussed 
in ATT using the 
unpublished data supplied by Suchitra Balachandran, of 11 member giants observed in IC 4651, only one had detectable Li with A(Li) = 1. The 
other 10 stars
only had upper limits, with the limits ranging from A(Li) = 1.05 to -0.1. \citet{pas04} observed 5 giants, three of which overlapped with 
the earlier 
sample. For the two new giants, the stars had upper limits to A(Li) near 0.2. For the stars in common to the surveys, the newer data showed 
an upper limit 
of 0.38 where the older sample found an upper limit of 0.2, a measured value of 0.5 where the older data implied an upper limit of 1.0, and 
a measured
value of 0.82 where the previous result was an upper limit of 0.75. Therefore, of 13 observed stars, one giant definitely has A(Li) at 1.0 with possibly a
second near 0.8. Eight more giants have a measured value of 0.5 or an upper limit of 0.7 or less. The remaining three giants have upper 
limits, but they
range between 0.95 and 1.05, so the possibility of detectable Li cannot be ruled out. If we class these last three as unknowns, the fraction of
giants with high Li (A(Li) $\sim$1) in IC 4651 is 2 out of 10, statistically the same as in NGC 752.

\section{SUMMARY}
The specific focus of this study has been the Li abundance distribution of stars at the turnoff of the intermediate-age open 
cluster, NGC 3680, with
the general goals of linking future Li analyses of turnoff stars in NGC 6253 to the larger array of published data in other open clusters while
reevaluating NGC 3680 itself in the context of the evolution of the Li-dip. After taking into account the modest differences 
in the reddening and temperature scales, there is very good agreement with past Li abundance determinations. The combined data set more than doubles the turnoff sample and increases the red
giant sample from 6 to 9 stars. The expanded sample better delineates the Li-dip region and reinforces the
observed dominance of the giant branch by stars with A(Li) near 1. 

Spectroscopic analysis of the red giants in NGC 3680 produces [Fe/H] = -0.17 $\pm$ 0.07 from 8 stars. Previous giant star spectroscopy gives the
same result if the spectra are analyzed with the same stellar parameters. The absolute scale can, however, be set at any value between [Fe/H]
= 0.00 and -0.30 by adopting microturbulence velocities between 1.3 km/sec and 1.9 km/sec. For the dwarfs, the scatter is a strong function of
the S/N; from 5 single stars at the turnoff with S/N above 110, [Fe/H] = -0.07 $\pm$ 0.02. There is no statistically significant evidence from
lines other than Fe that NGC 3680 has anything but scaled-solar elemental abundances. An average of all abundance determinations weighted by the 
inverse of the standard errors of the mean heavily favors the dwarfs and produces [Fe/H] = -0.08 $\pm$ 0.02 (sem). With the reddening and 
metallicity known, isochrone comparisons require an age of 1.75 Gyr and an apparent modulus of $(m-M)$ = 10.3. These data place the center of 
the Li-dip at a 1.35 $M_{\sun}$.

With the fundamental cluster information in hand, NGC 3680 is contrasted with two slightly younger clusters, NGC 752 and IC 4651, and the even
younger clusters, Hyades and Praesepe. Assuming a slightly higher reddening value of E$(B-V)$ = 0.12 for IC 4651, with [Fe/H] 
= -0.06 and +0.15 for NGC 752 and IC 4651, respectively, both clusters have an age near 1.5 Gyr, with apparent moduli of 8.42 
and 10.4, respectively.
For the Li-dip comparisons, the Hyades and Praesepe are both assumed to have zero reddening and [Fe/H] = +0.15; the distance moduli and any small
age difference between these two clusters are irrelevant for purposes of the discussion. The combined Hyades/Praesepe samples are used to
delineate a Li-dip profile, centered at $B-V$ = 0.44 or 1.42 $M_{\sun}$. Evolution of the profile to the mean age of the three older clusters
generates a CMD morphology that reproduces the observed Li-dip turnoff extremely well, implying that while the mass of stars in the center
of the dip varies positively with metallicity, 
the color profile of Li-depletion across the dip may be
similar but with the central color shifted to bluer values for 
lower [Fe/H]. Over the range from [Fe/H] = -0.1 to +0.15, the central mass
of the dip follows the relation M/M$_{\sun}$ = 1.38 ($\pm$0.04) + 0.4 ($\pm$0.2)[Fe/H]. Within the errors, there appears to be 
little change in the
degree of Li-depletion for stars within the dip for ages greater than the Hyades. For lower mass stars just below the Li-dip, there is some
evidence for modest depletion in Li ($\sim$0.2 dex) between 0.7 Gyr and 1.5 Gyr, assuming the clusters all formed with the same 
level of Li on the ZAMS.
For the giants, the color and Li dichotomy discussed in previous investigations has been confirmed and strengthened for NGC 752 and NGC 3680. 
The younger cluster has, relative to the first-ascent giant branch, a significantly bluer giant branch distribution than NGC 3680, while 
only 2 of
11 stars have detectable Li falling within the A(Li) = 1 bin, in contrast with 5 of 9 for NGC 3680. For IC 4651, the comparable 
numbers are 2 of 10,
as expected given its similarity to NGC 752, but these claims are tempered by the uncertainty in the true Li abundance for 3 additional giants.
If this pattern holds up, it strengthens the argument that, due to the lower mass of the stars leaving the main sequence, the giants 
populating the
supposed clump in NGC 3680 are first-ascent giants, placing an important constraint on the transition in evolved structure for stars 
in the mass range of 1.8 to 1.6 $M_{\sun}$.

The three clusters profiled in this investigation are regularly compared because they are so similar but also because they represent objects that,
by the standards of open cluster work, are well-studied. The spectroscopic sample of stars within NGC 3680 has been doubled by this analysis but
the Li-dip profile, as presented in Fig. 10, can only be adequately mapped through the superposition of data from all three clusters. 
One reason for the frequently qualified conclusions in the present paper, as well as others related to Li evolution, is the small number of single members, as is the case in NGC 3680; in other clusters, the paucity of data represents an inadequate fraction of cluster members incorporated in analyses.  As an example, the plateau effect for stars above the Li-dip seems apparent in Fig. 10, 
but delineation of the brighter stars' depletion as they evolve toward the giant branch is hampered by observations of only one star in this region
in NGC 752 and only five of more than two dozen possible candidates in IC 4651. A more complete set of observations has been obtained in
NGC 752 by one of the authors (CD), but an expanded sample in IC 4651, especially among the giants, could prove invaluable.

In the arena of future observations, two needs come to mind. One of the weaknesses of the metallicity-dependent  relation derived for the
mass of the center of the Li-dip is the small baseline in [Fe/H]. A critical test of its reliability could come from the observation of
a more metal-poor open cluster in a comparable age range. The obvious options are NGC 2420 and NGC 2506 with typical estimated abundances
\citep{taat, fri02, att06a, lee08} between [Fe/H] $= -0.3$ and $-0.4$, though, for a higher estimate for NGC 2506, see \citet{car04}. The second
somewhat surprising deficiency is the lack of modern, broad-band, wide-field CCD photometry for two of the three clusters discussed, IC 4651
and NGC 752. The $uvby$ survey of IC 4651 by \citet{mei00} is outstanding, but a complementary broad-band data base would make isochrone comparisons
more straightforward, particularly on the lower main sequence; the composite sample for NGC 752 clearly has a high but
uneven degree of precision and remains incomplete at fainter magnitudes.
 
\acknowledgements
The authors wish to thank the referee for thoughtful and constructive comments, and the staff at Cerro Tololo Inter-American Observatory for their invaluable assistance in obtaining the
observations that form the basis of this study. CTIO is operated by the Association of Universities for Research in Astronomy, under contract with
the National Science Foundation. Extensive use was made of the SIMBAD database, operating at CDS, Strasbourg, France and the WEBDA database 
maintained at the University of Vienna, Austria (http://www.univie.ac.at/webda). We also acknowledge with gratitude the use of IRAF, distributed
by the National Optical Astronomy Observatories.  CPD gratefully acknowledges support from
the National Science Foundation under grant AST-0607667. BJAT and BAT are also grateful to the Astronomy Department at Indiana University 
for hospitality during their Fall 2008 stay in Bloomington.

\clearpage
\figcaption[f1.eps]{Proper-motion probability distribution for stars identified as most likely to be potential members of NGC 3680. \label{f1}}

\figcaption[f2.eps]{Radial velocities for the stars in the current study as a function of the proper-motion probability. 
The dashed horizontal line defines the cluster mean velocity from the probable dwarf members, while the 
horizontal dash-dot lines define the radial-velocity limits for probable single-star cluster members, between $-1.8$ and 5.1 km/sec. \label{f2}}

\figcaption[f3.eps]{CMD for stars observed in the present study, excluding probable non-members. 
Filled circles are likely cluster members while open circles identify members that are suspected of being binaries. Triangles denote stars observed by PRP, with binaries similarly denoted by open symbols.
Superposed are Y$^{2}$ isochrones of appropriate metallicity with ages of 1.7 Gyr and 1.8 Gyr, adjusted for the reddening and apparent distance modulus 
of NGC 3680. \label{f3}}

\figcaption[f4.eps]{A(Li) vs. V for stars in NGC 3680. Filled circles and triangles are stars observed in this survey with detectable Li and upper
limits, respectively, combined with the data of PRP if the samples overlap. Open circles and triangles are stars with detectable Li and
upper limits, respectively, observed only by PRP but transferred to our temperature and Li scale. \label{f4}}

\figcaption[f5.eps]{CMD for NGC 752 superposed upon a Y$^{2}$ isochrones with [Fe/H] = -0.05, $(m-M)$ = 8.4,
and E$(B-V)$ = 0.03 and ages of 1.4 Gyr and 1.5 Gyr. Known binaries and non-members have been removed from the CMD. \label{f5}}

\figcaption[f6.eps]{The combined Li data for NGC 3680 (black symbols) and NGC 752 (red symbols). Circles are detected values while 
triangles are upper limits.  \label{f6}}

\figcaption[f7.eps]{CMD for IC 4651 superposed upon a 1.5 Gyr isochrone with [Fe/H] = +0.15, $(m-M)$ = 10.4,
and E$(B-V)$ = 0.12. Triangles are stars identified as spectroscopic binaries. Known non-members have
been removed from the CMD. \label{f7}}

\figcaption[f8.eps]{A(Li) as a function of $B-V$ for stars in the Hyades (blue symbols) and Praesepe (red symbols). Circles
are detectable values of Li while the triangles are upper limits. Solid line is an estimate of the mean
relation through the points in the region of the Li-dip. \label{f8}}

\figcaption[f9.eps]{A(Li) as a function of $V$ for stars in IC 4651. Circles are detectable measures while triangles are upper
limits. The solid line is the mean relation of Fig. 8 evolved to the age of IC 4651 and shifted in $V$ by 10.25 mag. \label{f9}}

\figcaption[f10.eps]{A(Li) as a function of $V$ for the combined cluster sample of Fig. 6 and Fig. 9. Blue circles are detectable
measures while triangles are upper limits for NGC 3680 and NGC 752; the red symbols have the same meaning for IC 4651. 
The solid line is the mean relation of Fig. 9. The solid relation and the data for IC 4651 have been shifted to fainter magnitudes 
by 0.09 mag in $V$. \label{f10}}
\newpage
\plotone{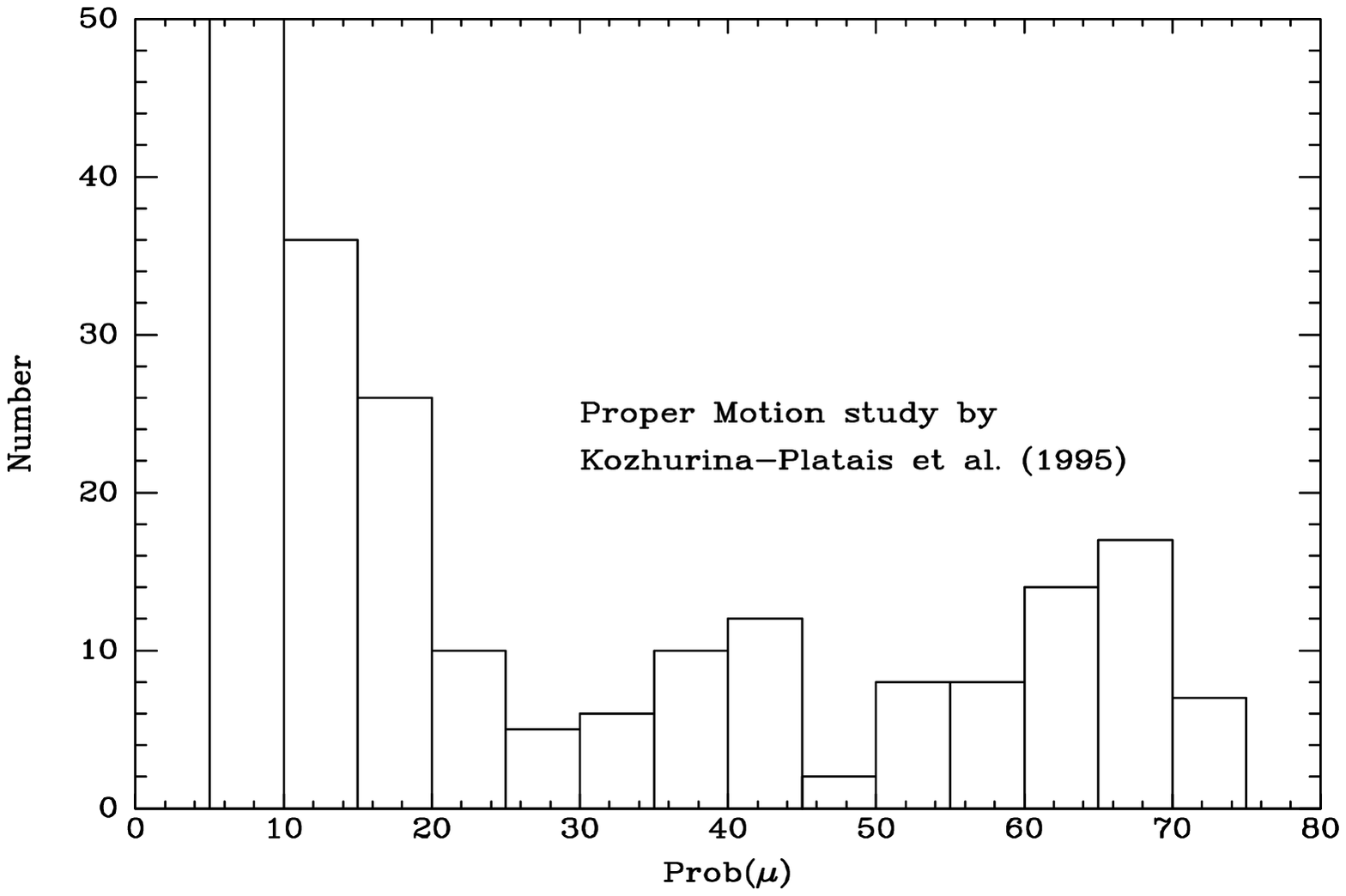}
\newpage
\plotone{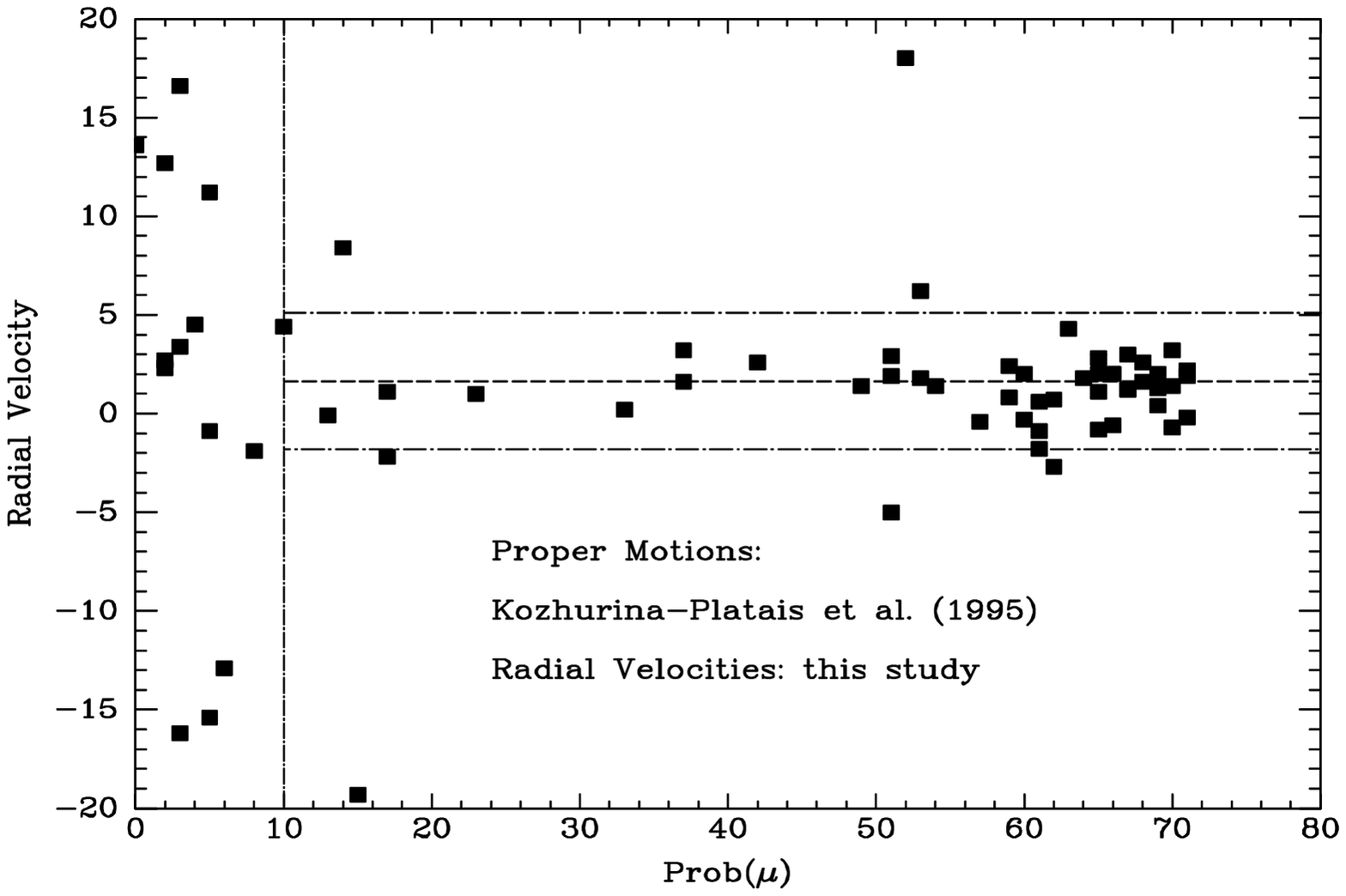}
\newpage
\plotone{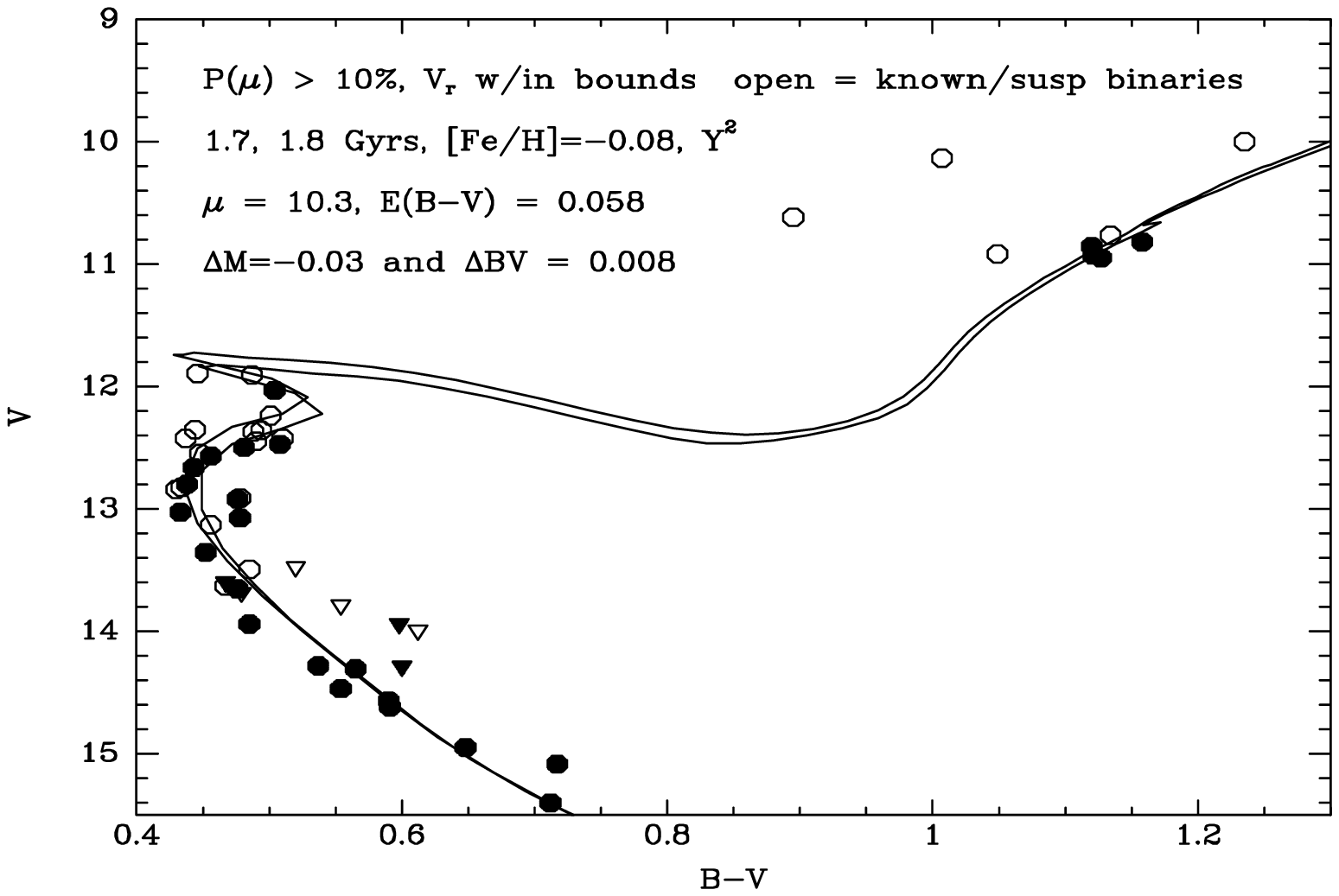}
\newpage
\plotone{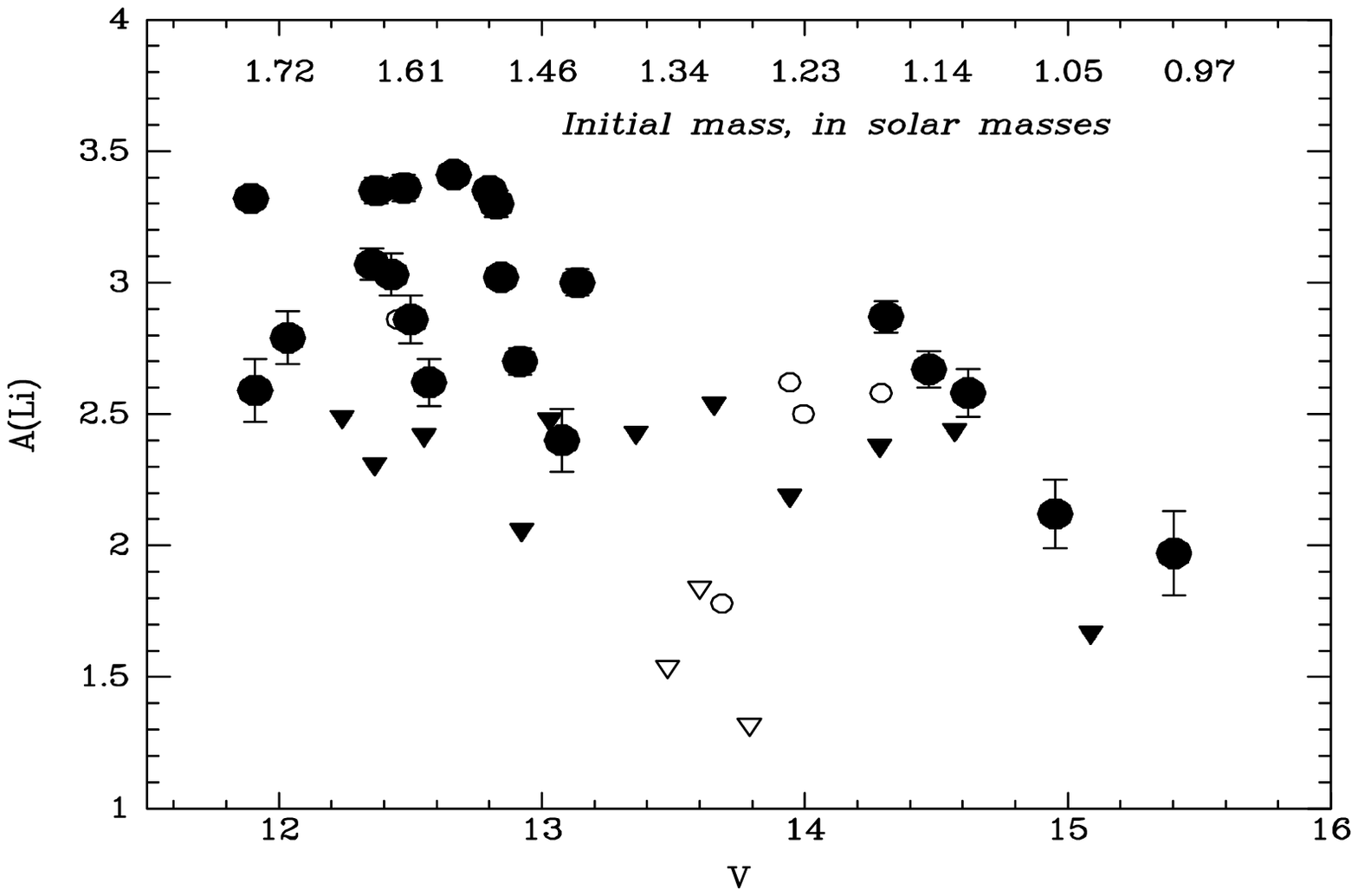}
\newpage
\plotone{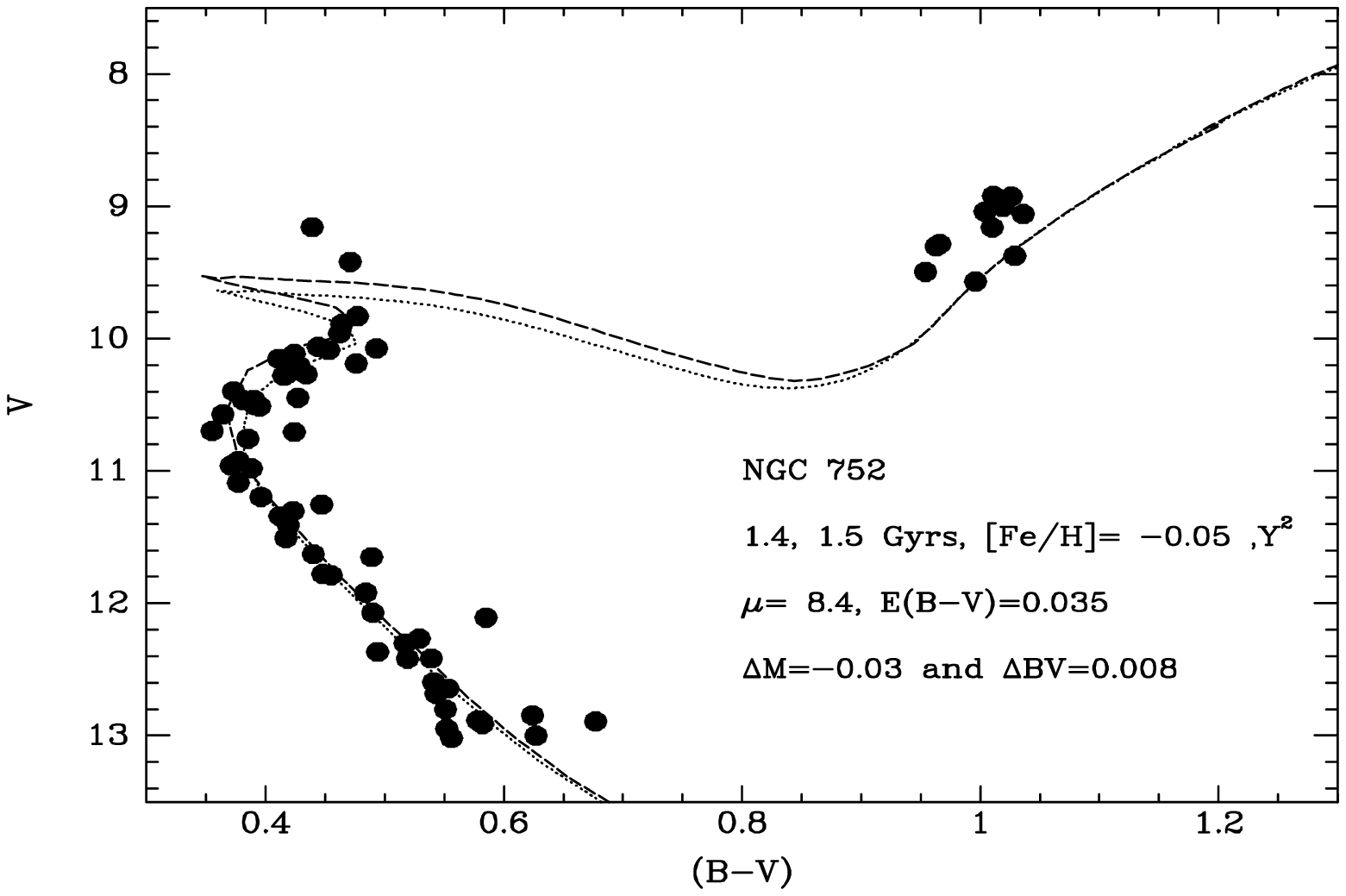}
\newpage
\includegraphics[scale=0.63,angle=-90]{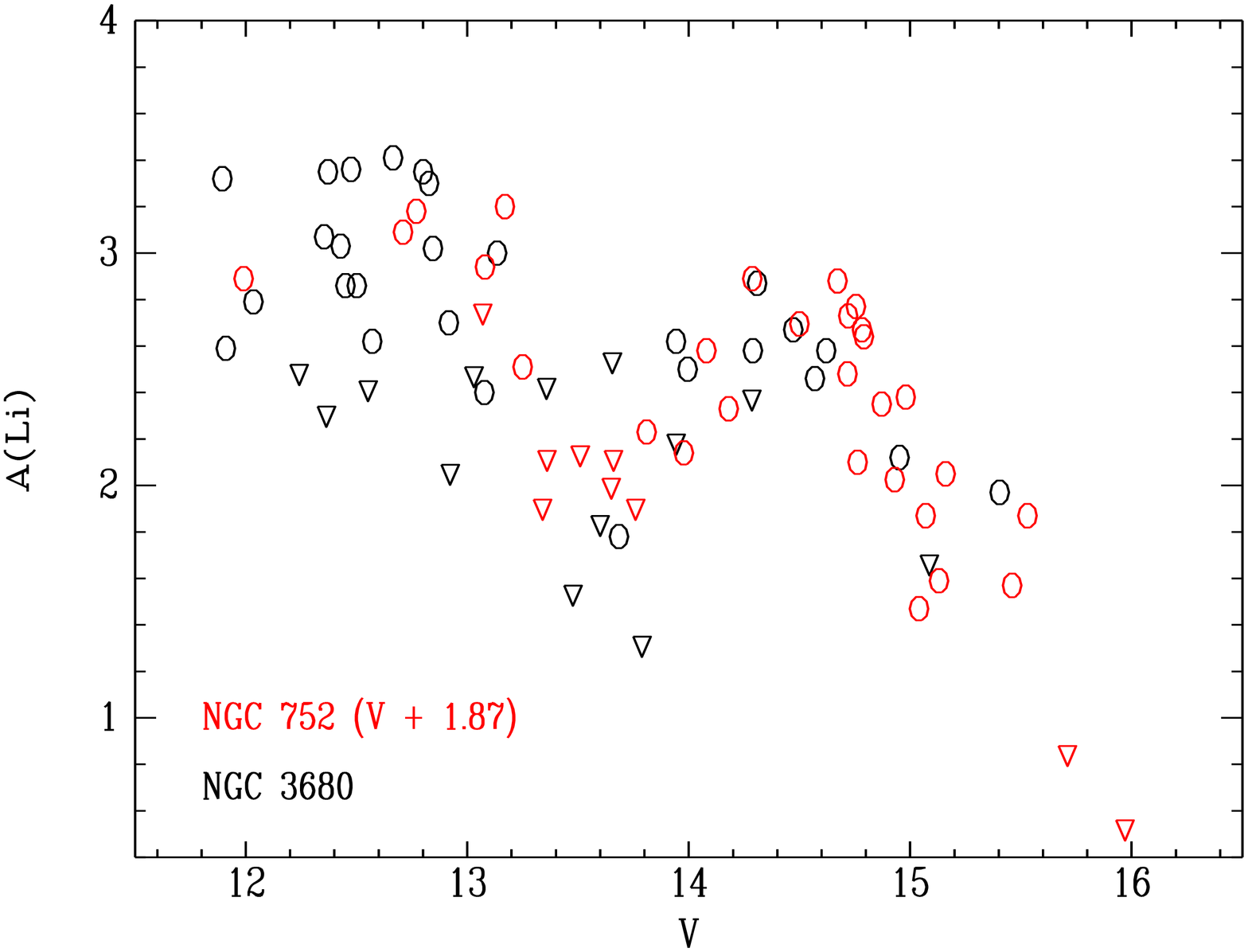}
\newpage
\plotone{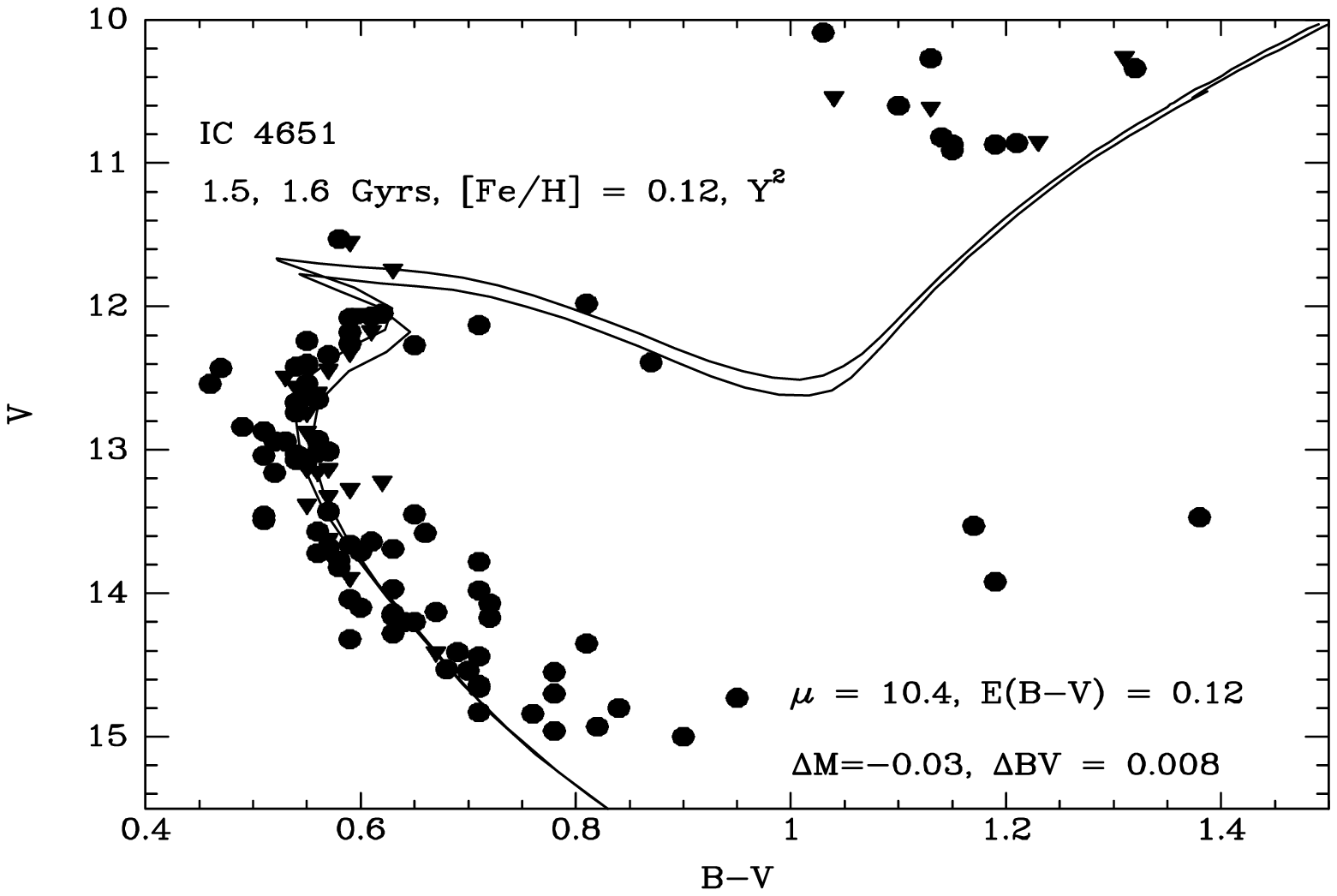}
\newpage
\includegraphics[scale=0.63,angle=-90]{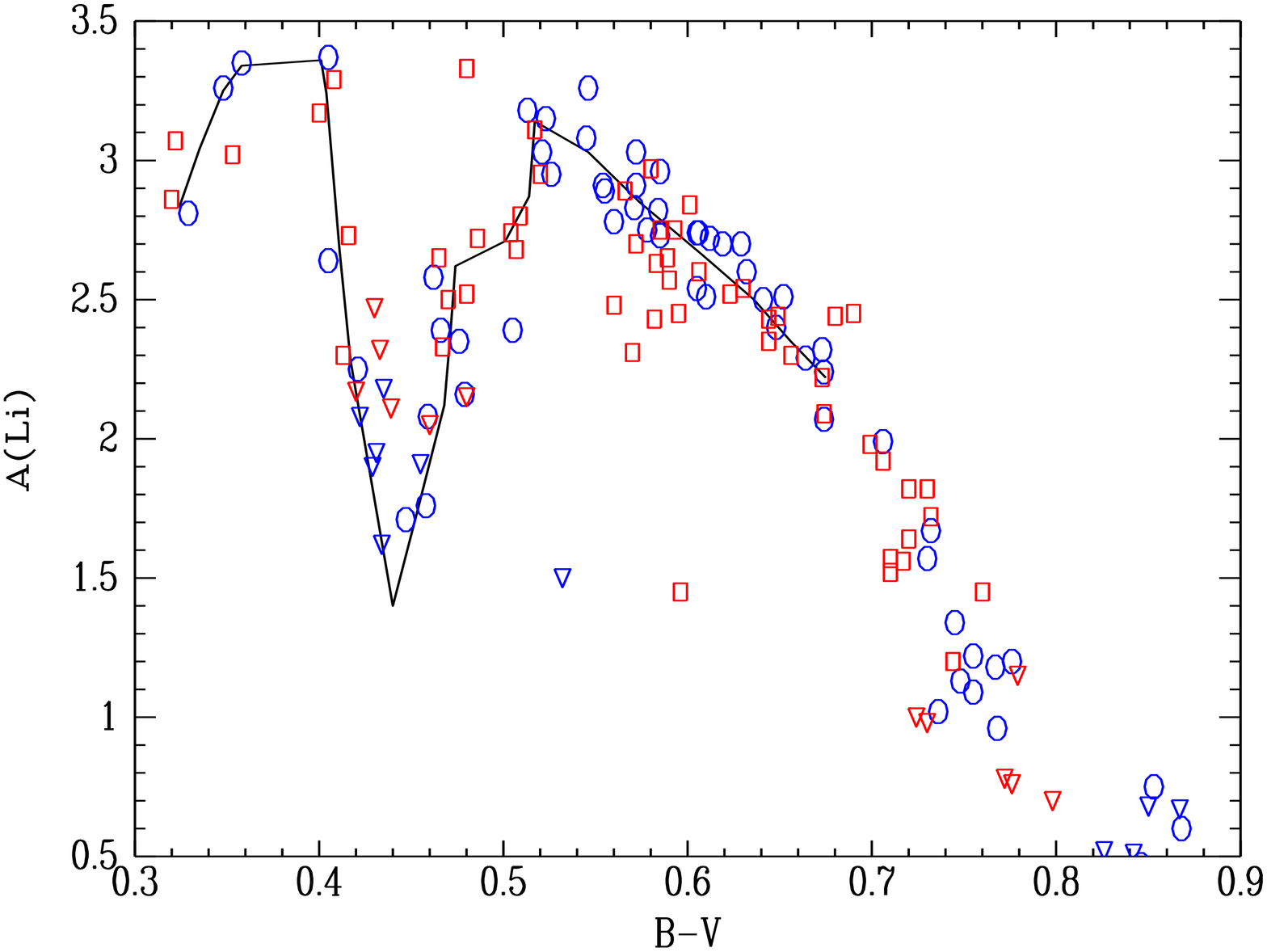}
\newpage
\plotone{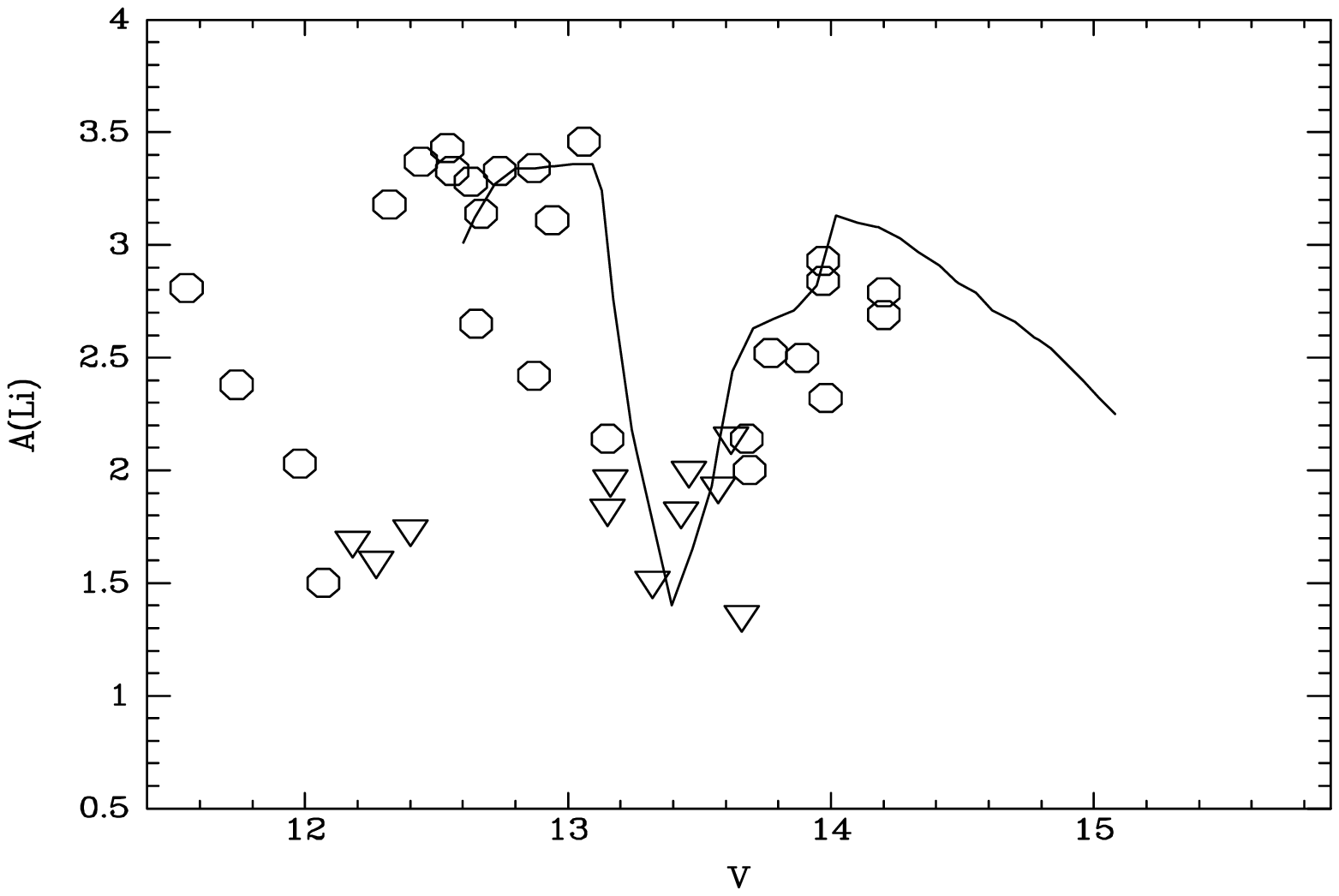}
\newpage
\includegraphics[scale=0.63,angle=-90]{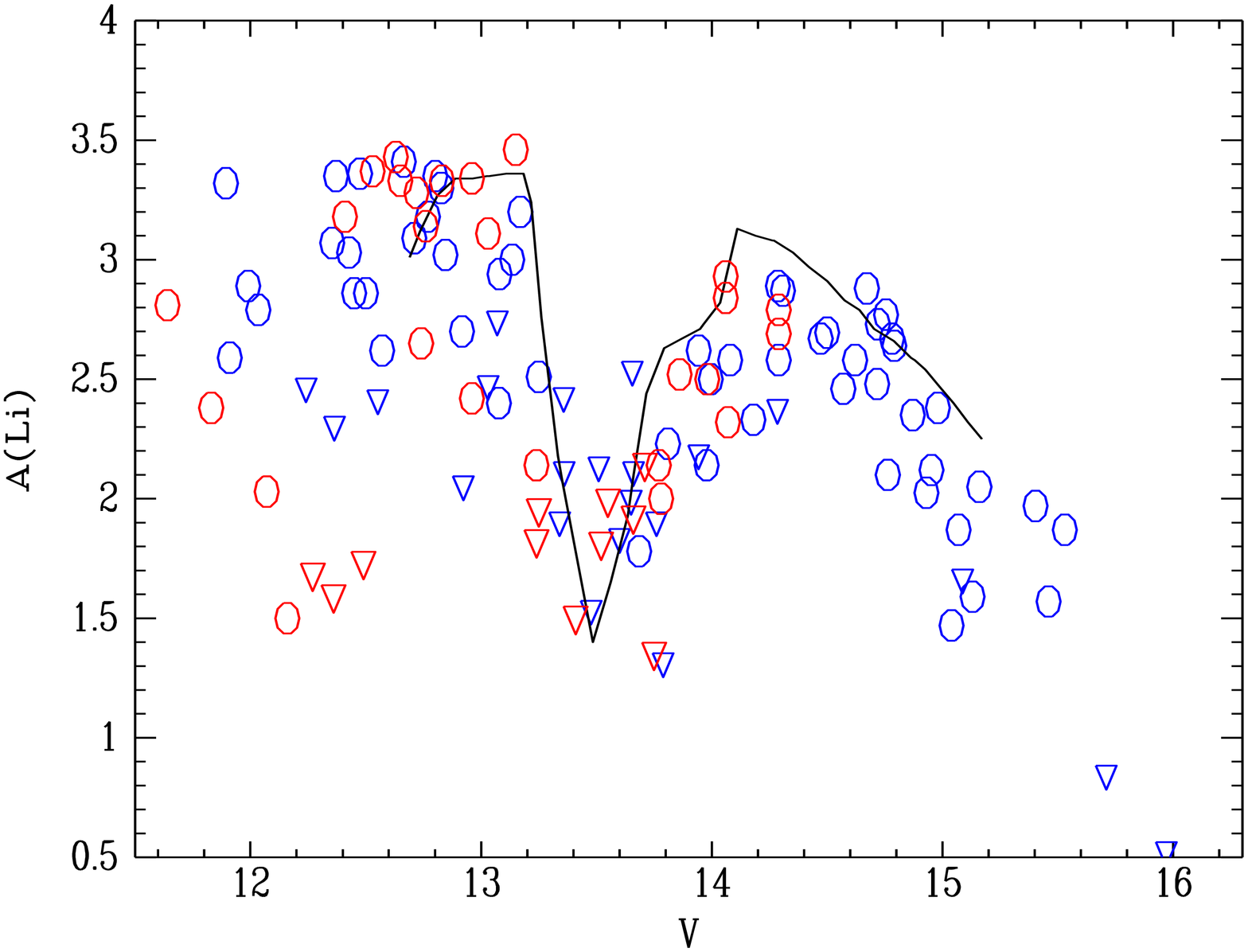}
\newpage
%\documentclass[preprint]{aastex}
%\begin{document}
\pagestyle{empty}
%\rotate
\begin{deluxetable}{cccrrrrrrrc}
\tablenum{1}
\scriptsize
\tablecolumns{11}
\tablewidth{0pc}
\tablecaption{Basic Data for Program Stars in NGC 3680} 
\tablehead{
\colhead{KGP}     & \colhead{WEBDA}     &
\colhead{Eggen}  & \colhead{$V$}     &
\colhead{$B-V$}     & \colhead{$P(\mu)$} &
\colhead{$V sin i$}     &
\colhead{$\sigma_{rot}$}     & \colhead{$V_{rad}$}     &
\colhead{$\sigma_{rad}$}  & \colhead{Comments}  }
\startdata
802  & 176  &   & 16.803 & 1.038 & 2  & 8.4 & 0.5 & 2.7 & \nodata & \cr
803  & 177  &   & 14.754 & 0.608 & 15 & 13.5 & 2.5 & -19.3 & 3.2 &\cr
833  & 183  &   & 15.520 & 0.773 & 5  & 13.4 & 0.5 & -0.9 & 1.0 &\cr
873  & 190  &   & 15.255 & 0.708 & 4  & 4.1 & 0.1 & 4.5 & 0.6  &\cr
936  & 4028 &   & 12.665 & 0.443 & 65 & 30.8 & 0.9 & 1.1 & 1.2 &\cr
945  & 3004 &   & 15.585 & 0.797 & 17 & 17.3 & 0.6 & -2.2 & 1.0 & sb?\cr
952  & 4081 &   & 12.371 & 0.488 & 71 & 14.3 & 0.4 & 1.9 & 0.9 & SB1(N97) \cr 
979  & 4119 &   & 16.756 & 1.028 & 3  & 14.7 & 0.6 & 16.6 & 1.2 & sb? \cr
983  & 4021 &   & 14.952 & 0.648 & 37 & 16.1 & 0.4 & 3.2 & 0.7 &\cr
988  & 79   & R & 11.827 & 0.451 & 51 & 58.4 & 2.4 & -5.0 & 3.1& w/sb?\cr
     &      & &  & & & & & & \\
1008 & 3095 &   & 12.241 & 0.501 & 64 & 40.6 & 1.8 & 1.8 & 2.3 & w/sb? \cr
1035 & 1    & 1 & 11.894 & 0.446 & 60 & 39.5 & 1.6 & -0.3 & 1.9 & SB2(N97) \cr
1050 & 70   & I & 14.621 & 0.591 & 37 & 10.6 & 0.3 & 1.6 & 0.7 &\cr
1083 & 4    & 4 & 12.923 & 0.476 & 61 & 24.1 & 0.6 & 0.6 & 0.9 &\cr
1097 & 3064 &   & 14.136 & 0.600 & 52 & 22.5 & 0.8 & 18.0 & 1.1 & SB? (KP97);sb?\cr 
1119 & 4073 &   & 16.333 & 0.926 & 3  & 7.6 & 0.2 & -16.2 & 0.8 &\cr
1123 & 4132 &   & 15.650 & 0.813 & 18 & 8.3 & 0.2 & 79.9 & 0.6 &\cr
1138 & 72   & K & 12.695 & 0.434 & 53 & 46.5 & 2.4 & 6.2 & 3.0 & w/sb?\cr
1168 & 14   & 14& 12.354 & 0.444 & 65 & 30.0 & 1.2 & -0.8 & 1.5 & SB2(N97) \cr
1175 & 13   & 13& 10.822 & 1.158 & 70 & 16.8 & 0.3 & 1.4 & 0.5 &\cr
     &      & &  & & & & & & \\
1192 & 15   & 15& 12.501 & 0.481 & 69 & 18.0 & 0.4 & 1.3 & 0.7 &\cr
1194 & 5    & 5 & 12.845 & 0.430 & 67  & 12.0 & 0.3 & 3.0 & 0.7 & SB1(N97) \cr
1197 & 16   & 16& 13.321 & 0.511 & 63 & 11.0 & 0.4 & 50.8 & 0.9 & SB2(N97)\cr
1202 & 4038 &  &  16.350 & 0.797 & 5 & 3.9 & 0.2 & -15.4 & 0.9 &\cr
1207 & 6    & 6 & 12.475 & 0.508 & 65 & 22.9 & 0.5 & 2.8 & 0.8 &\cr
1208 & 19   & 19& 12.364 & 0.494 & 70 & 45.4 & 3.9 & 3.2 & 4.8 & w/sb? \cr
1218 & 222  &  &  13.943 & 0.485 & 54 & 16.4 & 0.4 & 1.4 & 0.7 &\cr
1243 & 12   & 12& 15.041 & 0.705 &    & 12.1 & 0.2 & 13.6 & 0.6 &\cr
1261 & 20   & 20& 10.139 & 1.007 & 59 & 16.8 & 0.2 & 0.8 & 0.4 & SB2(N97,M95)\cr
1274 & 227  &  &  14.472 & 0.554 & 51 & 6.3 & 0.2 & 1.9 & 0.6 &\cr
    &       & &  & & & & & & \\
1286 & 11   & 11& 10.920 & 1.049 & 61 & 15.6 & 0.3 & -0.9 & 0.6 & SB1(N97,M95) \cr
1304 & 80   &21A& 14.762 & 0.724 & 2  & 21.5 & 1.1 & 2.3 & 1.7 & SB2(N97)\cr
1319 & 10   & 10& 12.827 & 0.434 & 71 & 40.6 & 1.5 & -0.2 & 1.7 & w/sb?\cr
1324 & 39   & 39& 12.428 & 0.510 & 66 & 11.9 & 0.3 & 2.0 & 0.7 & SB? (KP97)\cr
1331 & 81   &25A& 13.655 & 0.476 & 59 & 25.7 & 0.8 & 2.4 & 1.1 &\cr
\tablebreak
1347 & 38   & 38 &12.450 & 0.490 & 65 & 45.7 & 2.2 & 2.0 & 2.7 & SB1(N97)\cr
1350 & 33   & 33& 11.910 & 0.487 & 66 & 15.0 & 0.4 & -0.6 & 0.8 & SB1(N97)\cr
1352 & 3055 &   & 15.679 & 0.781 & 2  & 11.6 & 0.5 & 12.7 & 1.1 &\cr
1355 & 3081 &   & 16.582 & 0.866 & 5  & 10.6 & 0.5 & 11.2 & 1.3 &\cr
1365 & 3001 &   & 12.801 & 0.438 & 70  & 8.4 & 0.2 & -0.7 & 0.7 &\cr
1374 & 26   & 26& 10.952 & 1.127 & 62 & 17.5 & 0.3 & 0.7 & 0.5 &\cr
1376 & 35   & 35& 13.077 & 0.478 & 68 & 11.0 & 0.2 & 1.6 & 0.5 &\cr
1379 & 34   & 34& 10.623 & 0.895 & 63 & 19.6 & 0.4 & 4.3 & 0.6 & PhotBin(M95;AHTC)\cr
1395 & 37   & 37& 13.030 & 0.433 & 53 & 22.2 & 0.7 & 1.8 & 1.0 &\cr
1404 & 27   & 27& 10.770 & 1.134 & 71 & 16.0 & 0.3 & 2.2 & 0.6 & SB1O(M95,N97,M07)\cr
    &       & &  & & & & & & \\
1405 & 31   & 31& 13.135 & 0.456 & 69 & 21.6 & 0.6 & 0.4 & 0.8 & SB1(N97)\cr
1410A & 240B &  & 12.424 & 0.437 &\nodata & 60.3 & 3.1 & 1.1 & 4.1 & SB?(KP97); w/sb?\cr
1446 & 2058 &  &  15.801 & 0.795 & 16 & 10.6 & 0.4 & 51.2 & 0.9 &\cr
1454 & 32   & 32& 12.917 & 0.478 & 66 & 21.2 & 0.5 & 2.0 & 0.8 & SB1(N97)\cr
1461 & 41   & 41& 10.930 & 1.121 & 23 & 16.6 & 0.3 & 1.0 & 0.5 &\cr
1469 & 44   & 44& 10.002 & 1.235 & 67 & 17.2 & 0.3 & 1.2 & 0.6 & SB? (KP97)\cr
1506 & 42   & 42& 12.977 & 0.441 & 62 & 51.4 & 2.8 & -2.7 & 3.7 & sb?\cr
1507 & 46   & 46& 13.358 & 0.452 & 60 & 16.0 & 0.4 & 2.0 & 0.7 &\cr
1510 & 43   & 43 &12.552 & 0.448 & 68 & 30.6 & 1.6 & 2.6 & 1.9 & PhotBin(ATTS)\cr
1522 & 45   & 45& 12.034 & 0.504 & 67 & 30.6 & 1.1 & 1.2 & 1.4 &\cr
    &       & &  & & & & & & \\
1524 & 82   &46A& 13.498 & 0.485 & 49 & 29.3 & 0.8 & 1.4 & 1.0 & SB?(KP97) \cr
1534 & 1085 &  &  14.309 & 0.565 & 42 & 5.6 & 0.1 & 2.6 & 0.6 &\cr
1554 & 1124 &  &  14.285 & 0.537 & 51 & 11.0 & 0.3 & 2.9 & 0.8 &\cr
1575 & 2089 &  &  14.464 & 0.586 & 1  & 5.1 & 0.2 & -29.4 & 0.7 &\cr
1588 & 2005 &  &  16.079 & 0.852 & 8 & 6.6 & 0.3 & -1.9 & 1.1 &\cr
1610 & 2017 &  &  16.752 & 0.946 & 3  & 6.2 & 0.2 & 3.4 & 1.0 &\cr
1612 & 2084 &  &  15.404 & 0.712 & 13 & 13.5 & 0.3 & -0.1 & 0.5 &\cr
1624 & 2062 &  &  13.634 & 0.467 & 61 & 18.3 & 0.4 & -1.8 & 0.7 & SB1(N97)\cr
1627 & 2109 &  &  15.811 & 0.867 & 14 & 8.2 & 0.2 & 8.4 & 0.7 &\cr
1638 & 58   & 58& 12.891 & 0.434 & 58 & 21.1 & 0.8 & 22.7 & 1.1 & SB1(N97)\cr
    &      & &  & & & & & & \\
1649 & 2110 &  &  12.571 & 0.456 & 57 & 32.4 & 1.2 & -0.4 & 1.4 &\cr
1690 & 51   & 51 &(14.57) &(0.59)& 10 & 13.9 & 0.5 & 4.4 & 0.9 &\cr
1722 & 1019 &  &  15.087 & 0.717 & 33 & 17.5 & 0.3 & 0.2 & 0.6 &\cr
1815 & 1076 &  &  15.241 & 0.711 & 6  & 11.2 & 0.3 & -12.9 & 0.6 &\cr
1873 & 53   & 53& 10.859 & 1.120 & 69 & 18.0 & 0.3 & 2.0 & 0.7 &\cr
\enddata
\end{deluxetable}
%\end{document}

\newpage
%\documentclass[preprint]{aastex}
%\begin{document}
\pagestyle{empty}
%\rotate
\begin{deluxetable}{ccrrrrrrrcc}
\tablenum{2}
\tabletypesize\small 
%\scriptsize
\tablecolumns{9}
\tablewidth{0pc}
\tablecaption{Lithium EW and Abundance Data for Program Stars in NGC 3680} 
\tablehead{
\colhead{KGP}     & \colhead{Temp.}     &
\colhead{EW}  & \colhead{EW$'$} & \colhead{$\sigma_{EW}$}     &
\colhead{A(Li)}     & \colhead{$\sigma_{A}$} &
\colhead{S/N per pixel}     &
\colhead{FWHM}     & }
\startdata
\cutinhead{Abundances from detected Lithium Lines}
  873  &  5659  &  76.4  &  69.9 & 11.7 & 2.34 & 0.10  & 42 & 0.53 \cr
  936  &  6707  &  94.3  &  94.3  & 5.3 & 3.41 & 0.04 & 139 & 1.18 \cr
  952  &  6515  &  108.9 &  108.0 &  7.1 & 3.37 & 0.05  & 77 & 0.66 \cr
  983  &  5879  &  36.9  &  32.0 &  8.0 & 2.13 & 0.13  & 46 & 0.30 \cr
 1035  &  6694  &  84.1  &  84.1 &  5.4 & 3.33 & 0.04 & 132 & 1.10 \cr
 1050  &  6098  &  59.7  &  56.2  & 9.2 & 2.63 & 0.09  & 58 & 0.62 \cr
 1097  &  6063  &  70.4  &  66.7  & 6.6 & 2.70 & 0.06  & 84 & 0.68 \cr
 1138  &  6746  &  65.6  &  65.6  & 7.0 & 3.22 & 0.06 & 110 & 1.30 \cr
 1168  &  6703  &  54    &  54.0  & 6.3 & 3.08 & 0.06 & 107 & 0.99 \cr
 1175  &  4606  &  86.2  &  71.2 &  3.2 & 1.02 & 0.03 & 163 & 0.58 \cr
      &       &        &         &          &       &          &        &        \cr
 1192  &  6545  &  50.8  &  50.1 &  7.8 & 2.93 & 0.09 &  68 & 0.62 \cr
 1194  &  6763  &  46.1  &  46.1 &  3.5 & 3.03 & 0.04 & 133 & 0.47 \cr
 1197  &  6419  &  46.1  &  44.7 &  4.8 & 2.77 & 0.06 & 112 & 0.64 \cr
 1207  &  6432  &  120.7 &  119.3 &  7.4 & 3.38 & 0.05  & 89 & 0.95 \cr
 1261  &  4892  &  44.4  &  32.0 &  2.5 & 0.98 & 0.04 & 209 & 0.59 \cr
 1274  &  6244  &  51.6  & 49.1  & 6.6 & 2.68 & 0.07 &  66 & 0.42 \cr
 1286  &  4808  &  34.7  &  21.6  & 2.2 & 0.50 & 0.10 & 233 & 0.59 \cr
 1319  &  6746  &  70.6  &  70.6  & 5.4 & 3.26 & 0.05 & 124 & 0.99 \cr
 1324  &  6424  &  74.6  &  73.2  & 9.6 & 3.05 & 0.08 &  77 & 1.20 \cr
 1350  &  6519  &  28.7  &  27.9  & 6.7 & 2.61 & 0.12 &  69 & 0.47 \cr
      &       &        &         &          &       &          &        &        \cr
 1352  &  5404  &  104.8 &  96.5 & 24.9 & 2.26 & 0.17  & 21 & 0.60 \cr
 1365  &  6729  &  62.8  &  62.8 &  4.1 & 3.18 & 0.04 & 138 & 0.69 \cr
 1374  &  4662  &  92.2  &  77.8 & 10.2 & 1.15 & 0.08 &  56 & 0.72 \cr
 1376  &  6557  &  18.1  &  17.5 &  4.3 & 2.41 & 0.12 & 114 & 0.52 \cr
 1404  &  4649  &  85.7  &  71.2 &  2.3 & 1.09 & 0.02 & 225 & 0.61 \cr
 1405  &  6651  &  50.3  &  50.2 &  5.1 & 3.00 & 0.05 & 113 & 0.72 \cr
 1454  &  6557  &  42.6  &  42.0 &  4.4 & 2.84 & 0.06 & 136 & 0.80 \cr
 1461  &  4672   &  35.6 &  21.3  & 0.0 & 0.40 & 0.10 & 225 & 0.51  \cr
 1522  &  6448  &  47.3  & 46.0  & 8.6 & 2.81 & 0.10  & 74 & 0.88 \cr
 1534  &  6200  &  77.8  & 75.0  & 7.4 & 2.88 & 0.06  & 72 & 0.62 \cr
      &       &        &         &          &       &          &        &        \cr
 1612  &  5644  &  44.5  & 37.9 & 11.1 & 1.99 & 0.17  & 44 & 0.52 \cr
 1638  &  6746  &  35.4  & 35.4 &  4.5 & 2.89 & 0.06 & 109 & 0.52 \cr
 1649  &  6651  &  24.4  & 24.3 &  4.6 & 2.64 & 0.09 & 128 & 0.76 \cr
 1690  &  5857  &  66.1  & 61.0 &  9.9 & 2.46 & 0.09 &  48 & 0.50 \cr
 1873  &  4674  &  90.5  & 76.2 &  3.3 & 1.16 & 0.03 & 167 & 0.65 \cr
\tablebreak
\cutinhead{Upper Limit Lithium Abundances}
 1008  &  6461  &  20    &  & & 2.49 & & 55  & 0.42 \cr
 1083  &  6566  &  7.7   &  & & 2.06 & & 156 & 0.39 \cr
 1123  &  5297   &  9    &  & & 1.16 & & 49  & 0.21 \cr
 1208  &  6482   &  8    &  & & 2.31 & & 76  & 0.36 \cr
 1218  &  6528   &  7    &  & & 2.19 & & 80  & 0.20 \cr
 1243  &  5670   &  7.8  &  & & 1.77 & & 49  & 0.34 \cr
 1304  &  5602   &  $<$11  &  & & 1.88 & & 46  &  \cr
 1331  &  6566   &  $<$7   &  & & 2.54 & & 70 &  \cr
 1355  &  5126  &  36.2  &  & & 1.60 & & 25 & 0.50 \cr
 1379  &  5036  &  25.4  &  & & 0.82 & & 90 & 0.67 \cr
      &       &        &         &          &       &          &        &        \cr
 1395  &  6750  & $<$5    &  & & 2.48 & & 104 &  \cr
 1446  &  5357  &  9     &  & & 1.52 & & 39  & 0.34 \cr
 1469  &  4475   &  35.4 &  & & 0.00 & & 315 & 0.53  \cr
 1507  &  6668  &  10    &  & & 2.43 & &  82 & 0.39 \cr
 1510  &  6729  &  5     &  & & 2.42 & & 67  & 0.20 \cr
 1554  &  6313  &  25.8  &  & & 2.38 & & 62  & 0.54 \cr
 1575  &  6117  &  $<$9    &  & & 2.26 & & 59  &  \cr
 1627  &  5123  &  4     &  & & 1.13 & & 25  & 0.10 \cr
 1722  &  5627  &  19.6  &  & & 1.67 & & 60  & 0.43 \cr
 1815  &  5648  &  8     &  & & 1.79 & & 39  & 0.26 \cr
\enddata
\end{deluxetable}
%\end{document}

\newpage
%\documentclass[preprint]{aastex}
%\begin{document}
\pagestyle{empty}
%\rotate
\begin{deluxetable}{crrrrrrr}
\tablenum{3}
\tabletypesize\small
%\scriptsize
\tablecolumns{8}
\tablewidth{0pc}
\tablecaption{Metal Abundance Data for Program Stars in NGC 3680}
\tablehead{
\colhead{KGP}     & 
\colhead{[Fe/H]}  & \colhead{$\sigma_{fe}$} & 
\colhead{$N_{lines}$} & \colhead{[Si/H]} & 
\colhead{[Ca/H]}     & 
\colhead{[Ni/H]}     & \colhead{$\sigma_{Ni}$} 
}
\startdata
\cutinhead{Giants}
%                                           Ca  2 lines here 
1175  &      -0.13 & 0.13 & 8 &  0.20 &  \nodata &  -0.15 & 0.04 \cr
1261  &      -0.18 & 0.11 & 8 &  0.00 &  \nodata &  -0.33 & 0.05 \cr
1286  &      -0.22 & 0.06 & 8 & -0.16 &  \nodata &  -0.10 & 0.07 \cr
1374  &      -0.31 & 0.12 & 8 &  0.00 &  \nodata &  -0.21 & 0.20 \cr
1404  &      -0.10 & 0.13 & 8 &  0.13 &  \nodata &  -0.14 & 0.04 \cr
1461  &      -0.22 & 0.09 & 8 &  0.09 &  \nodata &  -0.21 & 0.06 \cr
1469  &      -0.14 & 0.16 & 8 &  0.17 &  \nodata &  -0.09 & 0.04 \cr
1873  &      -0.09 & 0.23 & 8 & -0.02 &  \nodata &  -0.10 & 0.11 \cr
\cutinhead{Dwarfs}
%                                  si-1    Ca 1 line
 952  &       0.00 & 0.22 & 6 &  0.02 & -0.05 & -0.08 & 0.26 \cr
{\bf 1083} & -0.09 & 0.22 & 6 & -0.20 & -0.32 & -0.35 & 0.37 \cr
{\bf 1194} & -0.08 & 0.17 & 6 & -0.13 & -0.14 & -0.11 & 0.15  \cr
1218  &       0.05 & 0.17 & 5 &  0.04 & -0.02 &  0.00 & 0.32 \cr
1324  &      -0.06 & 0.15 & 6 &  0.04 & -0.01 & -0.17 & 0.27 \cr
1331  &      -0.25 & 0.07 & 4 & -0.07 & -0.31 & -0.07 & 0.06 \cr
{\bf 1365} & -0.07 & 0.23 & 6 & -0.16 & -0.21 & -0.14 & 0.19 \cr
{\bf 1376} & -0.03 & 0.24 & 6 & -0.07 & -0.18 & -0.09 & 0.19 \cr
1395  &      -0.19 & 0.12 & 4 & -0.14 & -0.39 & -0.22 & 0.10  \cr
{\bf 1454} & -0.10 & 0.12 & 6 & -0.10 & -0.24 & -0.21 & 0.32 \cr
1507  &      -0.03 & 0.19 & 5 &  0.08 & -0.06 & 0.08  & 0.11 \cr
1534  &       0.14 & 0.30 & 6 &  0.07 & -0.13 & 0.01  & 0.20 \cr
1624  &       0.13 & 0.03 & 5 & -0.22 & -0.11 & -0.16 & 0.09 \cr
\enddata
\end{deluxetable}
%\end{document}

\newpage
%\documentclass[preprint]{aastex}
%\begin{document}
\pagestyle{empty}
%\rotate
\begin{deluxetable}{cccccc}
\tablenum{4}
\tabletypesize\small
%\scriptsize
\tablecolumns{6}
\tablewidth{0pc}
\tablecaption{Abundance Data by line for Program Stars in NGC 3680}
\tablehead{
\colhead{$\lambda$} & 
\colhead{[A/H]} &
\colhead{$\sigma$} &
\colhead{[A/H]} &
\colhead{$\sigma$} &
\colhead{$N_{stars}$} \\
\colhead{} & \multicolumn{2}{l}{Giants} & 
\multicolumn{2}{l}{Dwarfs} & \colhead{} \\
\cline{2-3} \cline{4-6} \\
}
\startdata
\cutinhead{Iron}
6581.2 & -0.23 &0.16 &       &     &   \cr
6608.0 & -0.25 &0.11 &       &     &   \cr
6609.1 &       &     & -0.23 &0.09 &11 \cr
6678.0 & -0.07 &0.10 & -0.20 &0.13 &13 \cr
6703.6 & -0.12 &0.13 &       &     &   \cr
6725.4 & -0.06 &0.15 &       &     &   \cr
6726.7 & -0.15 &0.13 & -0.03 &0.19 &13 \cr
6733.2 & -0.20 &0.08 &  0.20 &0.15 &10 \cr
6750.2 & -0.34 &0.09 &  0.01 &0.14 &13 \cr
6752.7 &       &     &  0.08 &0.10 &11 \cr
\cutinhead{Calcium}
6717.7 &       &     & -0.17 &0.12 &13 \cr
\cutinhead{Nickel}
6643.6 & -0.14 &0.11 & -0.27 &0.22 &13 \cr
6767.8 & -0.20 &0.11 &  0.03 &0.14 &13 \cr
6772.3 & -0.16 &0.11 & -0.11 &0.16 &13 \cr
\cutinhead{Silicon}
6721.8 &  0.05 &0.12 & -0.06 &0.10 &13 \cr
\enddata
\end{deluxetable}
%\end{document}

\end{document}